\documentclass[fp,twocolumn]{jpsj3}
\usepackage{txfonts}
\usepackage{subfigure}
\usepackage{graphicx}
\usepackage{mathabx}
\usepackage{pstricks}
\usepackage{wasysym}
\usepackage{color}
\usepackage{amssymb}
\definecolor{darkpurple}{rgb}{0.4, 0.0, 0.4}

\title{Ground state selection and spin--liquid behaviour in the classical Heisenberg
model on the breathing pyrochlore lattice}

\author{Owen Benton\thanks{owen.benton@oist.jp} and Nic Shannon} 
\inst{
Okinawa Institute of Science and Technology Graduate University, Onna-son,
Okinawa 904-0395, Japan
}

\title{Ground--state selection and spin--liquid behaviour in the classical Heisenberg
model on the breathing pyrochlore lattice}

\author{Owen Benton\thanks{owen.benton@oist.jp} and Nic Shannon} 
\inst{
Okinawa Institute of Science and Technology Graduate University, Onna-son,
Okinawa 904-0395, Japan
}

\abst{Magnetic pyrochlore oxides, including the spin ice materials,
have proved to be a 
rich field for the study of geometrical frustration in 3 dimensions.
Recently, a new family of magnetic oxides has been synthesised
in which the half of the tetrahedra in the pyrochlore lattice are inflated
relative to the other half, making an alternating array of small and
large tetrahedra.
These ``breathing pyrochlore'' materials such as
LiGaCr$_4$O$_8$, LiInCr$_4$O$_8$ and  Ba$_3$Yb$_2$Zn$_5$O$_{11}$
provide new opportunities in the study of frustrated 
magnetism. 
Here we provide an analytic theory for the ground state phase diagram 
and spin correlations for the minimal model of magnetism in breathing
pyrochlores: a classical nearest neighbour Heisenberg model with different exchange 
coefficients for the two species of tetrahedra.
We find that the phase diagram comprises a Coulombic spin liquid phase, 
a conventional ferromagnetic phase and an unusual antiferromagnetic
phase with lines of soft modes in reciprocal space, stabilised by an
order--by--disorder mechanism.
We obtain a theory of the spin correlations in this model using
the Self Consistent Gaussian Approximation (SCGA)
which enables us to discuss the development of correlations in  breathing pyrochlores
as a function of temperature, and we quantitatively characterise 
the thermal crossover from the limit of isolated tetrahedra to the strongly correlated limit
of the problem.
We compare the results of our analysis with the results of recent neutron scattering
experiments on LiInCr$_4$O$_8$.
}

\begin{document}


\maketitle

\section{Introduction}
\label{section:introduction}

Systems exhibiting geometrical
frustration have proved over recent years to be
fertile ground for the discovery of novel
emergent phenomena \cite{lee08, balents10}.
Amongst frustrated magnetic systems there
are now several known materials whose low
temperature physics displays novel collective
behaviour which cannot be understood using
the usual Landau picture, based on broken symmetries
\cite{castelnovo08, fennell09, yamashita09, han12, kimura13, hirschbirger15}.

At the forefront of the research interest in these systems
has been the study of magnetism on the pyrochlore lattice \cite{gardner10}
-a network of corner sharing tetrahedra.
Nearest neighbour antiferromagnetic coupling between
spins on this lattice fails to select a unique ordered state
and leads to an exponentially large ground state manifold
\cite{reimers91, moessner98-PRB58, moessner98-PRL80}.
This results in the formation of a classical spin liquid
with extensive residual entropy and algebraically
decaying spin correlations \cite{isakov04, henley10}.
The ground state manifold 
of this spin liquid
is defined by a flux conservation law
$\nabla \cdot {\bf B}=0$ and the excitations
which violate this law are  
deconfined ``charges'' which interact via an emergent Coulomb law.
For this reason, this spin liquid is known as a ``Coulombic spin liquid''.

\begin{figure}
\centering
\includegraphics[width=0.7\columnwidth]{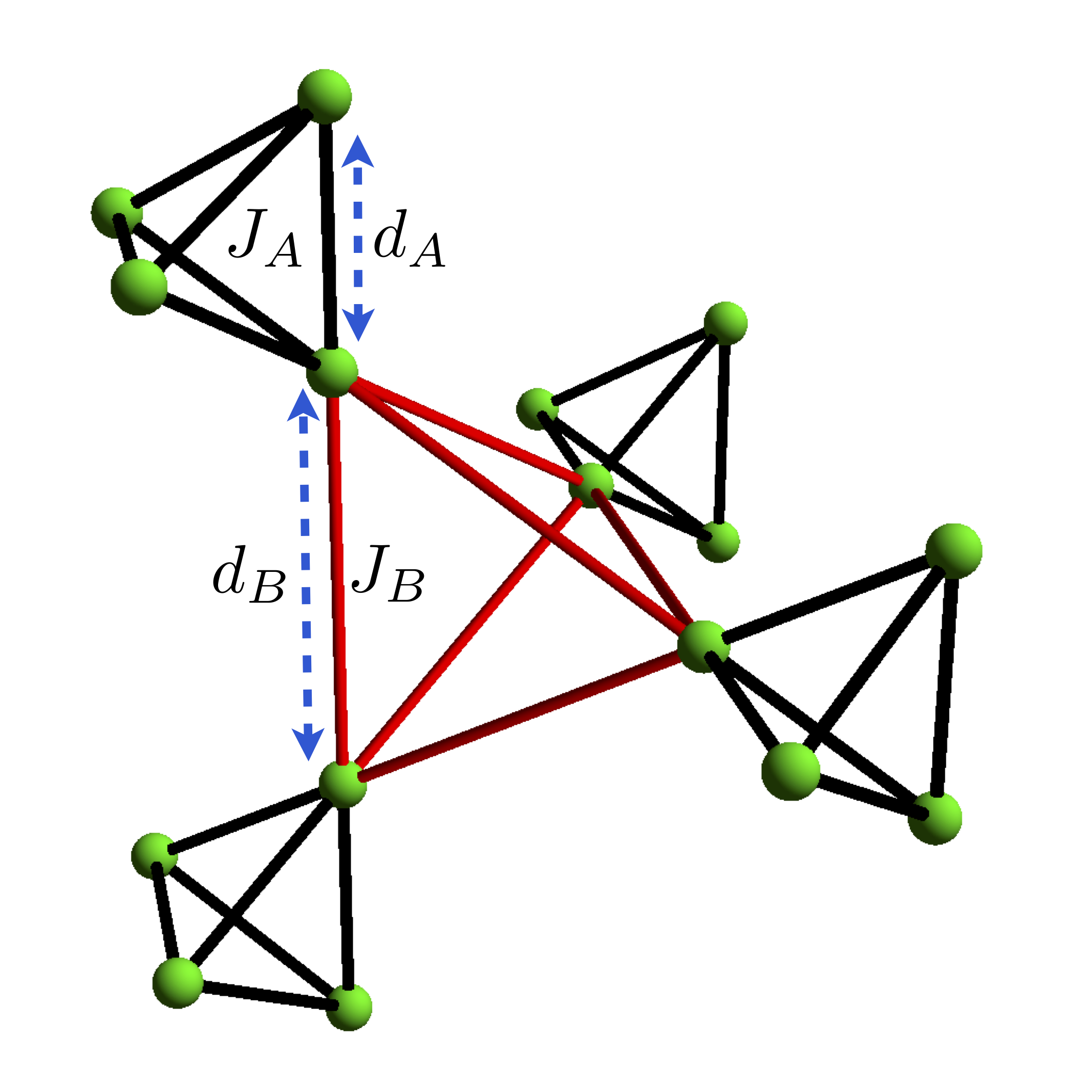}
\caption{(Color online)
The breathing pyrochlore lattice.
Like the pyrochlore lattice,
the breathing pyrochlore lattice is composed of
corner-sharing tetrahedra.
However, in the breathing pyrochlore lattice these
tetrahedra alternate in size.
The spins reside at the vertices of this lattice.
The two species of tetrahedra (here labelled `A' and `B', colored black and red)
have different bond lengths $d_A$ and $d_B$ and different exchange coefficients
$J_A$ and $J_B$.
 }
\label{fig:lattice}
\end{figure}

The classical Coulomb spin liquid finds realization in several experimental 
systems.
The spin ice materials 
R$_2$M$_2$O$_7$ (R=Ho, Dy; M=Ti, Sn) at low temperature
are famous examples of Coulomb liquids of Ising spins \cite{castelnovo12}.
Meanwhile, the chromium spinels MCr$_2$O$_4$ (M=Zn, Cd, Hg)
can be understood
in terms of a Coulomb liquid of vector spins interacting with fluctuations of 
the lattice.
As a consequence these materials have emerged as canonical examples
of the phenomenon of ``order-by-distortion'' in which magnetic frustration
is relieved by a structural transition \cite{yamashita00, tchernyshyov02, ueda06, ji09}.
Calculations based on a classical effective model for the spins with coupling
to the lattice predict  a rich phase diagram
in an applied magnetic field \cite{penc04, shannon06, motome06, penc07, shannon10}
which was observed in a series of remarkable experiments at fields up to 600T
\cite{ueda05, kojima08, miyata11-PRL107, miyata11-JPSJ80, miyata12}.

Recently, a new class of magnetic materials has been synthesised
which promises an interesting variation on this picture
\cite{okamoto13, tanaka14, kimura14, okamoto15}.
In the spinels LiInCr$_4$O$_8$
 and LiGaCr$_4$O$_8$
\cite{okamoto13, tanaka14, okamoto15, nilsen15-1}
 the Cr$^{3+}$
ions, carrying spin $S=\frac{3}{2}$, reside on a ``breathing'' pyrochlore
lattice depicted in Fig. \ref{fig:lattice}.
In this lattice the tetrahedra of the pyrochlore lattice are alternately small
and large, with the consequence that the two sets of tetrahedra have
different exchange coefficients.
Similarly, in the effective $S=\frac{1}{2}$ magnet, 
Ba$_3$Yb$_2$Zn$_5$O$_{11}$, the magnetic Yb$^{3+}$ ions
reside on a breathing pyrochlore lattice \cite{kimura14}.

As with Cr spinels, these oxide materials have antiferromagnetic
nearest-neighbour exchange interactions \cite{okamoto13}.
However this interaction has contributions from both direct and
indirect exchange, with opposite sign, and is very sensitive small
changes in bond angles or distances \cite{yaresko07, yaresko08} .
As a result, there is a very real possibility that breathing
pyrochlores based on larger anions, such as S or Se, might have
ferromagnetic interactons, in at least one of the the two sets of
tetrahedra.

Theoretical studies of the pyrochlore magnets with
unequal exchange interactions on the two species of tetrahedra
have concentrated on antiferromagnetic interactions, in the
quantum limit $S = 1/2$, as a route to understanding the 
$S = 1/2$ Heisenberg model on a pyrochlore lattice 
\cite{harris91, canals98, canals00, tsunetsugu01}. 
Performing perturbation theory about the limit of independent 
independent tetrahedra, Harris {\it et al.} \cite{harris91}
and Tsunetsugu \cite{tsunetsugu01} have explored
the possibility of a novel dimerized ground state.
Meanwhile Canals and Lacroix \cite{canals98, canals00}, 
considered the possibility of a quantum spin liquid.
At present, there is no established study of the Heisenberg model
on the breathing pyrochlore lattice in the presence of ferromagnetic 
interactions, or treatment of spin correlations within the 
high-temperature paramagnetic phase.
The recent synthesis of breathing pyrochlore magnets 
makes this a pressing issue.

In this article we provide a comprehensive
treatment of the ground state phase diagram and spin correlations
of the nearest neighbour
Heisenberg model on the breathing pyochlore lattice, for all
signs and ratios of exchange couplings.
The classical treatment should be particularly relevant for the
relatively large--spin $S=\frac{3}{2}$, Cr$^{3+}$ systems,
in the temperature regime above the structural transitions 
which are observed in  LiInCr$_4$O$_8$ at  15.9 K
and LiGaCr$_4$O$_8$ at 13.8 K \cite{okamoto13, tanaka14, okamoto15, nilsen15-1}.

We find that the nature of the classical ground state depends only the signs
of the two exchange coefficients associated with the two 
species of tetrahedra, not on their relative magitudes.
Where both exchange coefficients 
are antiferromagnetic 
the system enters a Coulomb phase at low temperature.
Where both exchange coefficients 
are ferromagnetic 
the system orders into a simple ferromagnetic phase.
Where the exchange coefficients differ in sign, the classical
ground state manifold
has a degeneracy of $\mathcal{O}(L)$,
where $L$ is the linear dimension of the system, 
with antiferromagnetic $[001]$
planes becoming effectively decoupled from one another.
This degneracy is then susceptible to being lifted by fluctuations, stabilising antiferromagnetic
order, meaning that this region of the phase diagram combines the
properties of dimensional reduction and order-by-disorder noted 
in other pyrochlore magnets \cite{ross09, ross11-PRB84, savary12-PRL109,  
zhitomirsky12, yan-arXiv}.
By using the Self Consistent Gaussian Approximation (SCGA) 
\cite{canals01, pickles08, conlon09, conlon10} we are then able to calculate the
behaviour of the spin correlations as a function of the exchange parameters
and temperature.

The article is structured as follows:

In Section \ref{section:phasediagram}
we obtain the phase diagram of the classical Heisenberg model
on the breathing pyrochlore lattice.


%

In 
Section \ref{section:ferromagnet}
we discuss the case of ferromagnetic coupling. 
We use the SCGA to show the temperature development of the spin correlations
and quantify the thermal crossover from isolated tetrahedra
to long range correlations by
 deriving equations for the
temperature dependence of the ferromagnetic correlation length.

In Section \ref{section:orderbydisorder} we discuss the case of 
exchange parameters with differing signs.
We show that the degenerate zero-energy modes which occur
in this case lead to a ``square ring'' structure observable in the
neutron scattering response.

In Section \ref{section:coulombphase} we discuss the case of
antiferromagnetic coupling which leads to a Coulombic spin liquid at
low temperature.
We show how the algebraic correlations of the spin liquid emerge
out of the short range, single tetrahedron, correlations with
decreasing temperature.

In Section \ref{section:discussion} we conclude with a summary of our results and a
comparison of our predictions with the results of neutron scattering 
experiments on the breathng pyrochlore  LiInCr$_4$O$_8$ \cite{okamoto15}.

\section{Ground state phase diagram}
\label{section:phasediagram}

\begin{figure}
\centering
\includegraphics[width=0.8\columnwidth]{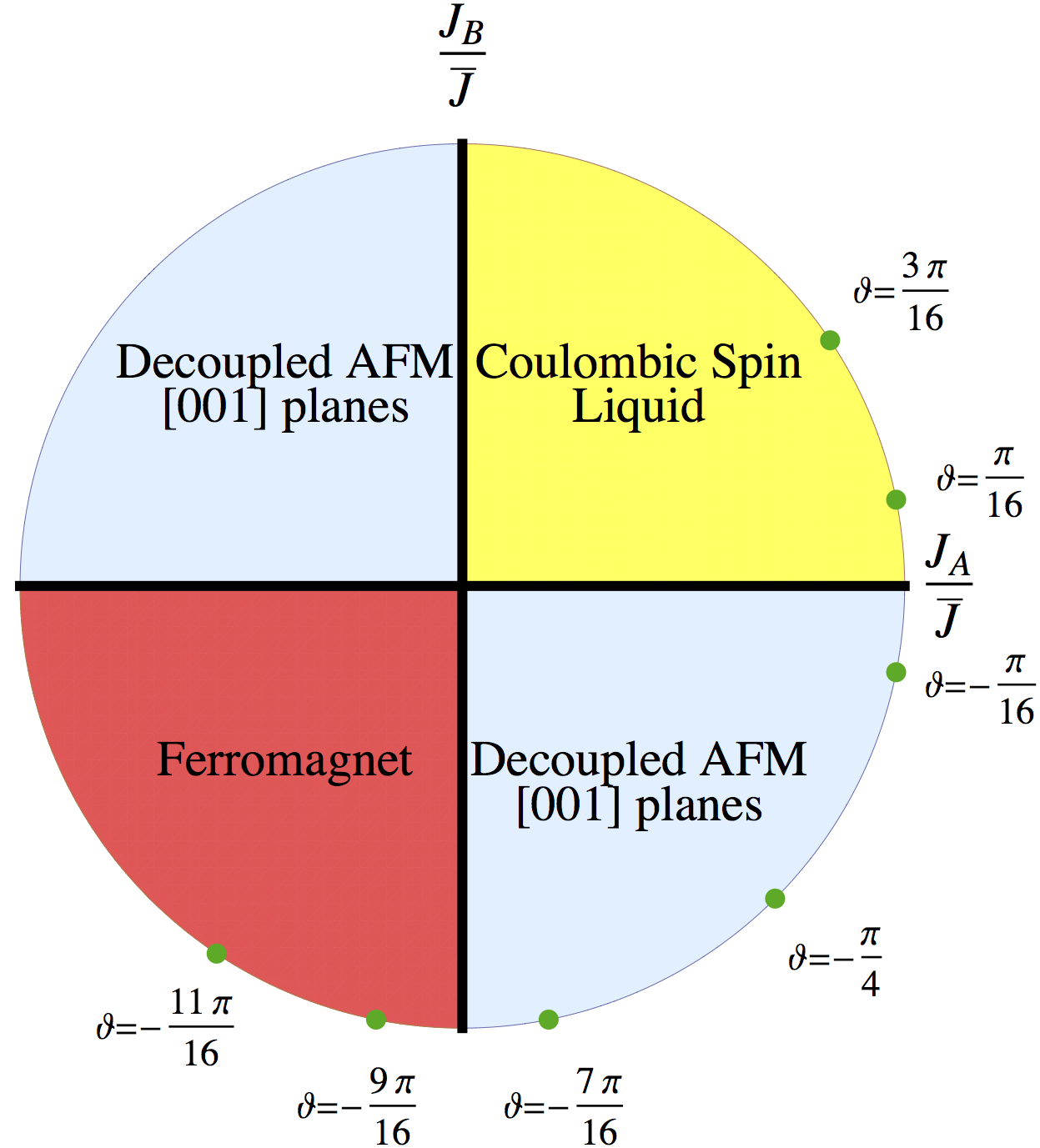}
\caption{(Color online)
Classical ground state phase diagram of the
breathing pyrochlore lattice Heisenberg model [Eq. (\ref{eq:Hbreathing})].
The ground state manifold is determined by the
signs of the
two couplings $J_A$ and $J_B$.
Where both couplings are ferromagnetic than the ground state is a simple collinear
ferromagnet.
Where both couplings are antiferromagnetic the system enters a macroscopically
degnerate Coulomb phase, in which the ground state manifold contains
all states where every tetrahedron has vanishing magnetisation.
If $J_A$ and $J_B$ have opposite signs than the set of classical ground states
is described by decoupled $[001]$ planes and has a degeneracy $\mathcal{O}(L)$.
This degeneracy will be lifted by the order--by--disorder mechanism.
Due to a correspondence with Heisenberg model on the FCC lattice [Eq. (\ref{eq:HFCC})]
we expect this to favour antiferromagnetic states with ordering vector ${\bf q}_{\sf ord}=(1, 0, 0)$
[Fig. \ref{fig:FCC-AFM}].
The green points indicate the parameter sets for which we plot
the spin correlation functions in Figs. \ref{fig:FMSq}-\ref{fig:coulombpowder}.
}
\label{fig:phasediagram}
\end{figure}

We consider the Heisenberg Hamiltonian
on the breathing pyrochlore lattice
\begin{eqnarray}
\mathcal{H}_{\sf breathing}=J_A \sum_{\langle ij \rangle_A} {\bf S}_i \cdot {\bf S}_j
+J_B \sum_{\langle ij \rangle_B} {\bf S}_i \cdot {\bf S}_j 
\label{eq:Hbreathing}
\end{eqnarray}
The two terms in Eq. (\ref{eq:Hbreathing}) correspond to the two sets
of tetrahedra on the pyrochlore lattice, which we label $A$ and $B$ \mbox{[cf. Fig.~\ref{fig:lattice}]}. 
Every bond of the lattice belongs uniquely to one tetrahedron and is
associated with an exchange coupling $J_A$ or $J_B$ depending
on which set of tetrahedra it belongs to, as illustrated in Fig. \ref{fig:lattice}.

The classical ground state phase diagram 
of Eq. (\ref{eq:Hbreathing}) is shown in Fig. \ref{fig:phasediagram}.
As is commonplace for two parameter models, we can re-express 
the exchange coefficients in terms of an overall energy scale $\bar{J}$
and an angle $\vartheta$:
\begin{eqnarray}
J_A=\bar{J} \cos(\vartheta), \ \ J_B=\bar{J} \sin(\vartheta)
\label{eq:thetadef}
\end{eqnarray}
There are three distinct phases on the phase diagram:
\begin{enumerate}
\item{For $0<\vartheta<\frac{\pi}{2}$ ($J_A>0, J_B>0$) the ground state
manifold contains all states where the  magnetisation
of every tetrahedron in the lattice vanishes and the system is
in a Coulomb phase at low temperature.}
\item{For $-\pi<\vartheta<-\frac{\pi}{2}$ ($J_A<0, J_B<0$) the 
ground state is a collinear ferromagnetic state 
}
\item{For $-\frac{\pi}{2}<\vartheta<0$ ($J_A>0, J_B<0$) 
or $\frac{\pi}{2}<\vartheta<\pi$ ($J_A<0, J_B>0$)  the 
ground state is comprises an $\mathcal{O}(L)$ set
of antiferromagnetic states with wavevectors 
of the form of ${\bf q}_{\delta}=(\delta, 0, 1)$
(and those related by symmetry).
This ground state manifold may be mapped directly
to the classical ground state manifold of the nearest
neighbour FCC antiferromagnet \cite{henley87, gvozdikova05}, 
as we shall show below.
}
\end{enumerate}

The phase diagram shown in Fig. \ref{fig:phasediagram}
may be derived using the following argument.
As in the case of the Heisenberg model on the ordinary (non-breathing)
pyrochlore lattice, the Hamiltonian [Eq. (\ref{eq:Hbreathing})] can be 
rewritten as a function only of the magnetisation of each tetrahedron.
We define, for a tetrahedron $t$, the magnetisation density
\begin{eqnarray}
{\bf m}_t=\frac{1}{4S}
\sum_{i \in t} {\bf S}_i
\end{eqnarray}
where the sum runs over the four spins at the vertices of tetrahedron $t$.
Eq. (\ref{eq:Hbreathing}) can then be rewritten solely in terms 
of the variables ${\bf m}_t$
\begin{eqnarray}
&&\mathcal{H}_{\sf breathing}= 8 J_A S^2 \sum_{t \in A} {\bf m}_t^2
+8 J_B S^2 \sum_{t \in B} {\bf m}_t^2
\nonumber \\
&&\qquad \qquad \qquad \qquad 
 - \frac{(J_A+J_B)}{2} N S^2  
\label{eq:heisenbergsimplification}
\end{eqnarray}

It is clear from Eq. (\ref{eq:heisenbergsimplification}) that when
$J_A$ and $J_B$ are both antiferromagnetic, any state where
\begin{eqnarray}
{\bf m}_t =
\begin{pmatrix}
0 \\
0 \\
0
\end{pmatrix}
 \qquad \forall \ t
\end{eqnarray}
is a classical ground state. 
This constraint gives rise to a Coulomb phase \cite{henley10}.

It is similarly clear from Eq. (\ref{eq:heisenbergsimplification}) that when
$J_A$ and $J_B$ are both ferromagnetic, the ground state is obtained
when every tetrahedron has a maximal value of ${\bf m}_t^2$
\begin{eqnarray}
{\bf m}_{t}^2=1 \qquad \forall \ t
\end{eqnarray}
This condition is ony satisfied by collinear ferromagnetic states.

Where the signs of $J_A$ and $J_B$ differ we can see from Eq. (\ref{eq:heisenbergsimplification})
that the condition to be in a ground state is that all ferromagnetic tetrahedra
have aligned spins but the intervening antiferromagnetic tetrahedra have vanishing
magnetisation.
Taking, without loss of generality
$$
J_A>0, \ \ J_B<0
$$
we require
\begin{eqnarray}
{\bf m}_t^2=0 \forall \ t \ \in \ A
\label{eq:AFMconstraintB}
\\
{\bf m}_t^2=1 \forall \ t \ \in \ B
\label{eq:FMconstraintA}
\end{eqnarray}

\begin{figure}
\centering
\subfigure[The FCC lattice as a network of edge-sharing
tetrahedra]{%
\includegraphics[width=.28\textwidth]{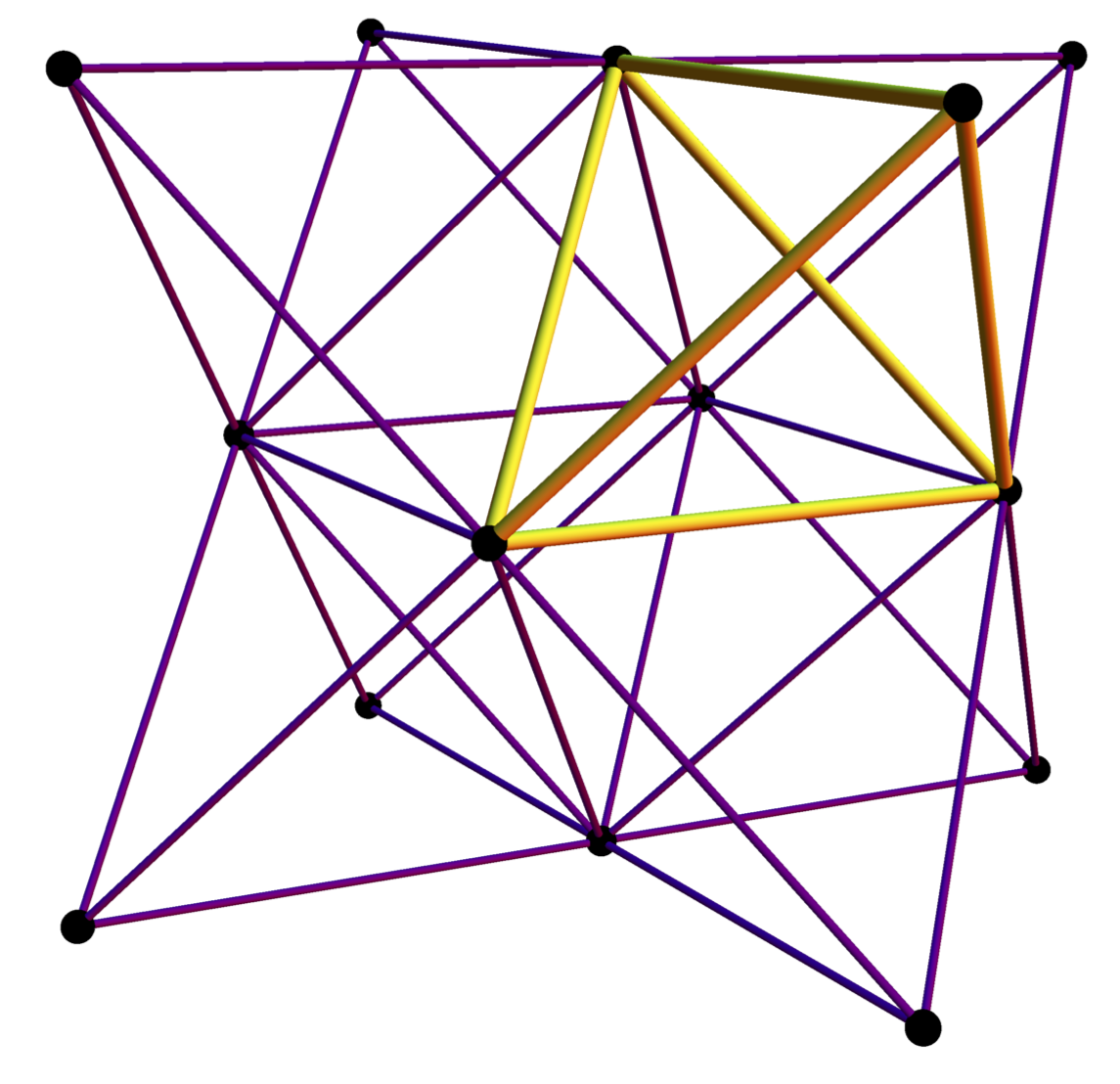}}%
\\ 
\subfigure[The $(100)$ state on the FCC lattice ]{%
\includegraphics[width=.28\textwidth]{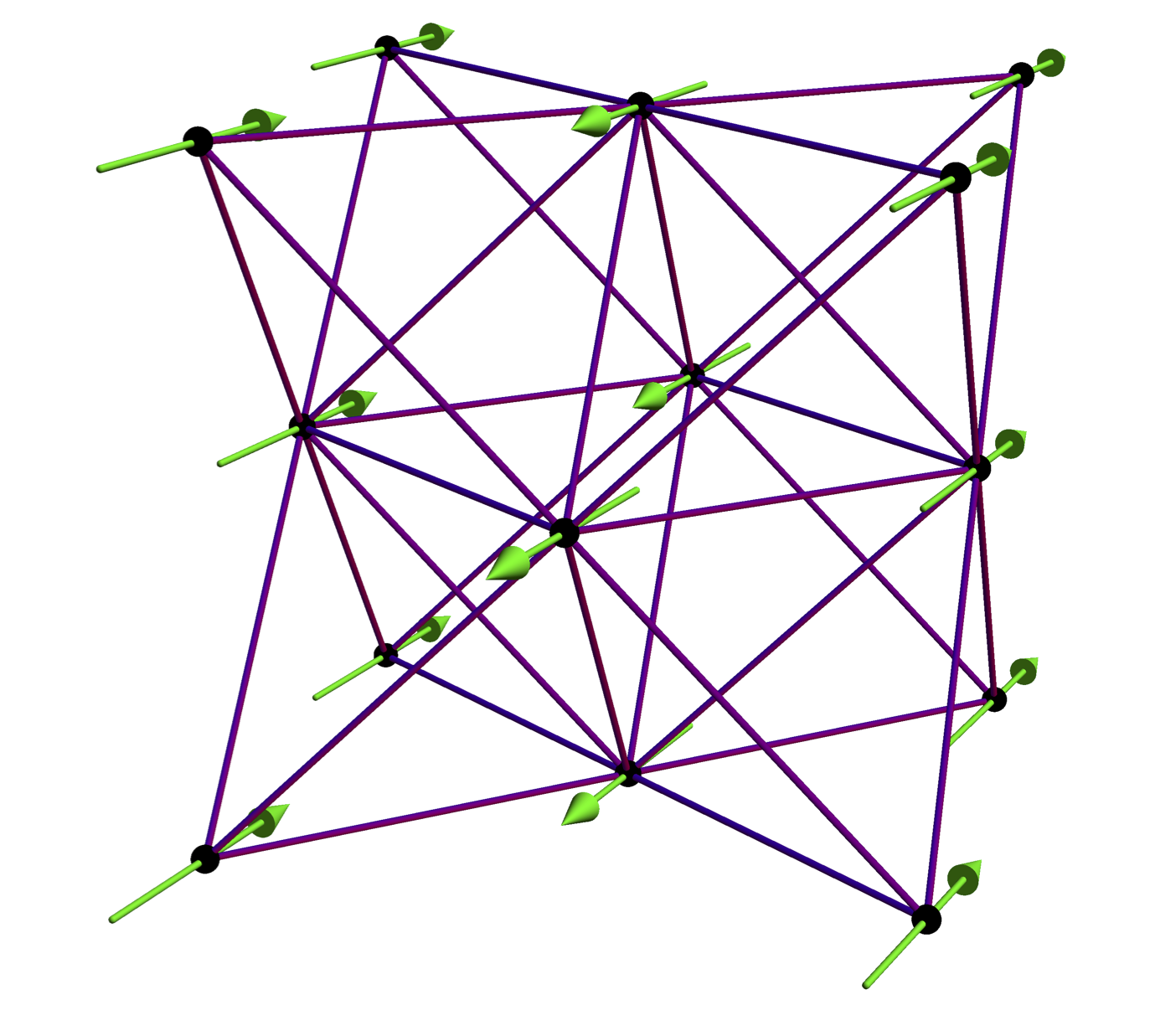}}%
\\
\subfigure[The $(100)$ state on the breathing pyrochlore lattice ]{%
\includegraphics[width=.28\textwidth]{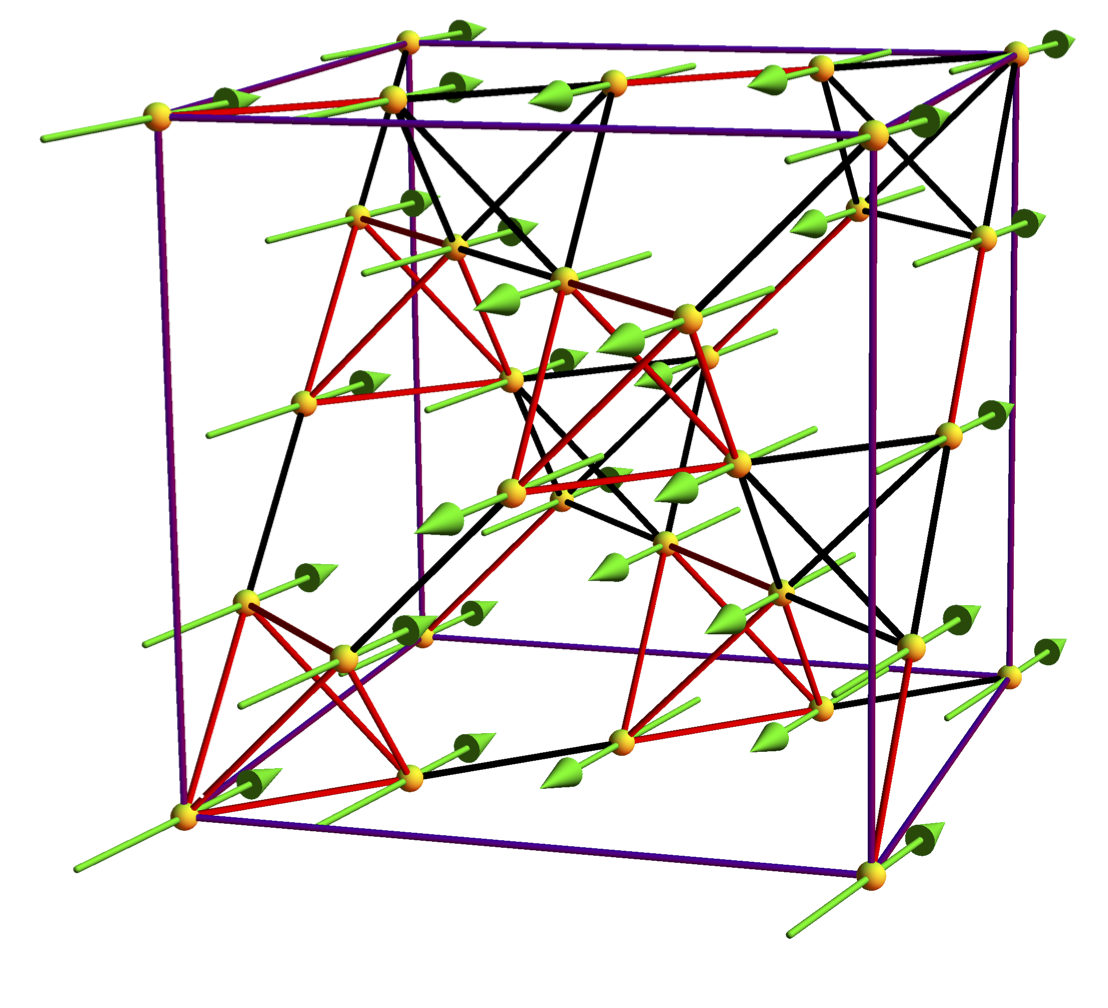}}%
\caption{(Color online). 
The mapping to the FCC antiferromagnet when $J_A$ and $J_B$
have opposite signs and the resulting antiferromagnetic order.
Each sublattice of tetrahedra ($A$ and $B$) of the breathing pyrochlore
lattice forms an FCC lattice, which can be viewed as a network of
edge sharing tetrahedra, as seen in $A$.
Where the exchange coefficients $J_A$ and $J_B$ differ in sign, the
classical ground state manifold of Eq. (\ref{eq:Hbreathing}) is the same
as the ground state manifold of the nearest neighbour antiferromagnet on
the FCC lattice, with each ferromagnetic tetrahedron of the breathing pyrochlore
model corresponding to a single spin of the FCC model.
This model has an $\mathcal{O}(L)$ set of ground states with fluctuations favouring collinear states
with ordering wavevector ${\bf q}_{\sf ord}=(1, 0, 0)$ (and states related by global symmetries).
As a result, the FCC model orders into the configuration shown in (b).
Since the two models have an exactly corresponding structure of soft modes,
the order by disorder mechanism should favour the analagous state on the breathing
pyrochlore lattice [shown in (c)] when $J_A$ and $J_B$ have opposite signs.
 }
\label{fig:FCC-AFM}
\end{figure}

These constraints lead to an equivalent classical
ground state manifold to the nearest neighbour
Heisenberg model on the FCC lattice.
We can see this by considering the Hamiltonian [Eq. (\ref{eq:Hbreathing})]
restricted to the subspace where Eq. (\ref{eq:FMconstraintA})
is perfectly obeyed. 
Since this fixes the exchange energy of all the $B$
tetrahedra, the Hamiltonian within this set of configurations
becomes
\begin{eqnarray}
\mathcal{H}_{\sf breathing}^{\sf FM-B}=
8 J_A S^2 \sum_{t \in A} {\bf m}_t^2 + \frac{(3J_B-J_A)}{2} N S^2 
\label{eq:HFM-A}
\end{eqnarray}.

Since all the spins on each $B$ tetrahedron are ferromagnetically aligned
the magnetisation of the $A$ tetrahedra is determined uniquely
by the total magnetisation of the four surrounding $B$ tetrahedra
\begin{eqnarray}
{\bf m}_{t_A}= \frac{1}{4}\sum_{t_B \in {\sf n.n.} t_A} {\bf m}_{t_B}.
\label{eq:neighbourmag}
\end{eqnarray}
Inserting Eq. (\ref{eq:neighbourmag}) into Eq. (\ref{eq:HFM-A})
and using Eq. (\ref{eq:FMconstraintA})
we obtain a nearest neighbour antiferromagnetic interaction between
$A$ tetrahedra.
\begin{eqnarray}
\mathcal{H}_{\sf breathing}^{\sf FM-B}=
 J_A S^2 \sum_{\langle t_B t_B' \rangle} {\bf m}_{t_B} \cdot {\bf m}_{ t_B'} + \frac{3J_B}{2} N S^2 
\label{eq:HFCC}
\end{eqnarray}
where the sum $\sum_{\langle t_B t_B' \rangle}$
runs over nearest neighbours of the FCC lattice of $B$
tetrahedra.

The FCC lattice may be depicted as a network of {\it edge-sharing}
tetrahedra as shown in Fig. \ref{fig:FCC-AFM}(a).
A ground state for the nearest neighbour antiferromagnet on this lattice
is obtained when each of these edge-sharing tetrahedra have vanshing
magnetisation.
Under the mapping from the breathing pyrochlore model [Eq. (\ref{eq:heisenbergsimplification})]
to the FCC model [Eq. (\ref{eq:HFCC})] this requirement is equivalent to Eq. (\ref{eq:AFMconstraintB}).

One ground state configuration of Eq. (\ref{eq:HFCC}) is depicted in Fig. \ref{fig:FCC-AFM}(b).
This is an antiferromagnetic state with ordering wavevector ${\bf q_{\sf ord}}=(1, 0, 0)$.
The corresponding state on the breathing pyrochlore lattice is depicted in Fig. \ref{fig:FCC-AFM}(c)
and one can readily verify that it satisfies Eqs.(\ref{eq:AFMconstraintB})-(\ref{eq:FMconstraintA}).

Further ground states can be generated from the configuration in Fig. \ref{fig:FCC-AFM}(b), 
by taking (e.g.) $[001]$ planes of spins and rotating them with respect to the rest of the lattice.
This operation keeps the system in the classical ground state manifold.
By generating ground states using these operations we see that the ground state manifold
contains states with wavevectors
\begin{eqnarray}
{\bf q}_{\delta}=(1, 0, \delta)
\label{eq:qdelta}
\end{eqnarray}
 (and those wavevectors related to Eq. (\ref{eq:qdelta}) by lattice symmetries).

From these considerations we can see that states with wavevectors
\begin{eqnarray}
{\bf q}_{\sf ord}=(1, 0, 0)
\label{eq:qord}
\end{eqnarray}
are special because they live at the junctions of the ground state
manifold, where two lines of classical ground states
(in the case of Eq. (\ref{eq:qord})  those with ${\bf q}_{\delta}=(1, 0, \delta)$
and \mbox{${\bf q}_{\delta}=(1, \delta, 0)$}) meet.
The resulting freedom to fluctuate for zero energy cost should favour these
states for selection via the order--by--disorder mechanism \cite{moessner98-PRB58, yan-arXiv}.

In the nearest neighbour FCC model it is known that fluctuations do indeed
favour  states with
 ordering wavevector ${\bf q}_{\sf ord}=(1, 0, 0)$ (referred to in the literature as Type I AFM states)
\cite{henley87, gvozdikova05}.
This entropic preference for states with \mbox{${\bf q}_{\sf ord}=(1, 0, 0)$}
 results in an
first order phase transition into these states 
\cite{gvozdikova05}.

Since the ground state manifold of the breathing pyrochlore model [Eq. (\ref{eq:Hbreathing})] with
$J_A<0, \ J_B>0$ (or equivalently $J_A>0, \ J_B<0$) is the same as in the FCC model
[Eq. (\ref{eq:HFCC})], the
structure of the soft modes is also the same, and it therefore seems reasonable to postulate
that the ground state selection will also be the same.

In the limit $|J_A|>>|J_B|$, for $J_A<0, \ J_B>0$, i.e. where the ferromagnetic exchange
is much stronger than the antiferromagnetic, fluctuations which are internal to the ferromagnetic
tetrahedra will cost a lot of energy and [Eq. (\ref{eq:HFCC})] will be a faithful treatment not only of
the ground state manifold but also of the low energy fluctuations. 
Therefore, in this limit at least, the ground state selection must be the same as in the FCC model.
Since tuning away from the limit $|J_A|>>|J_B|$ does not change the structure of the
ground state manifold we make the conjecture that the ${\bf q}_{\sf ord}=(1, 0, 0)$
state is selected across the whole region where $J_A$ and $J_B$ have opposite signs,
although this would need to be confirmed by simulation.

In the temperature range above the phase transition, two planes neighbouring $[001]$
planes with short range order at ${\bf q}=(1, 0, 0)$ will become effectively decoupled and
fluctuate essentially independently.
This gives rise to an apparent dimensional reduction, which will manifest itself as line like
features in the structure factor in the paramagnetic phase, in a similar phenomenon to that
observed in the paramagnetic phase of the rare earth pyrochlore Yb$_2$Ti$_2$O$_7$
\cite{ross09, ross11-PRB84, yan-arXiv}.
We will discuss this point further where we calculate the spin correlations in this region
of the phase diagram in Section \ref{section:orderbydisorder}.

\section{$J_A<0, \ J_B<0$: ferromagnetic phase}
\label{section:ferromagnet}

\begin{figure*}[t]
\centering
\includegraphics[width=.8\textwidth]{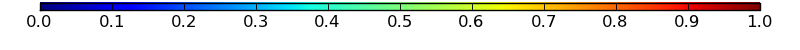}
\\
\subfigure[$\frac{T}{\bar{J}}=5, \vartheta=\frac{-11 \pi}{16}$]{%
\includegraphics[width=.4\textwidth]{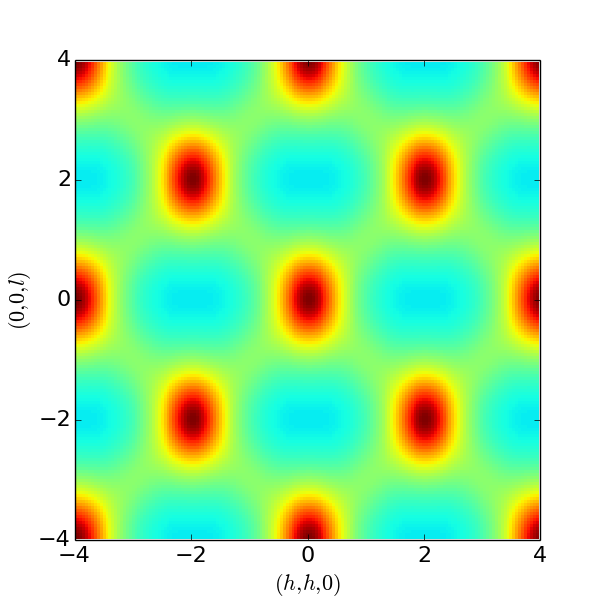}}%
\subfigure[$\frac{T}{\bar{J}}=5, \vartheta=\frac{-9 \pi}{16}$]{%
\includegraphics[width=.4\textwidth]{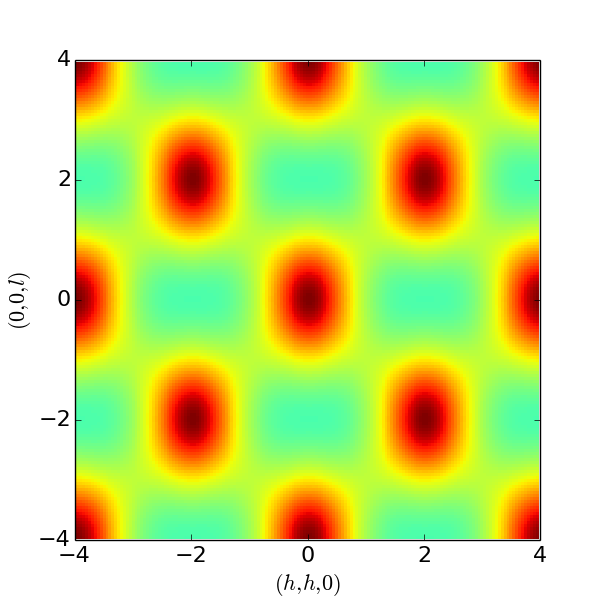}}%
\\
\subfigure[$\frac{T}{\bar{J}}=2, \vartheta=\frac{-11 \pi}{16}$]{%
\includegraphics[width=.4\textwidth]{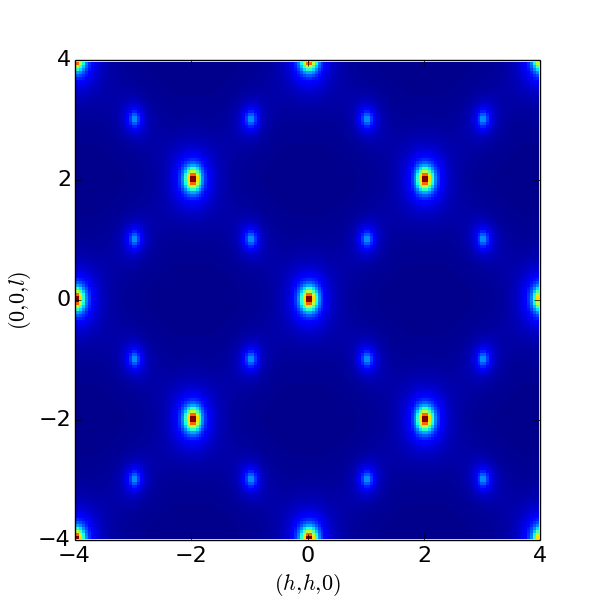}}%
\subfigure[$\frac{T}{\bar{J}}=2, \vartheta=\frac{-9 \pi}{16}$]{%
\includegraphics[width=.4\textwidth]{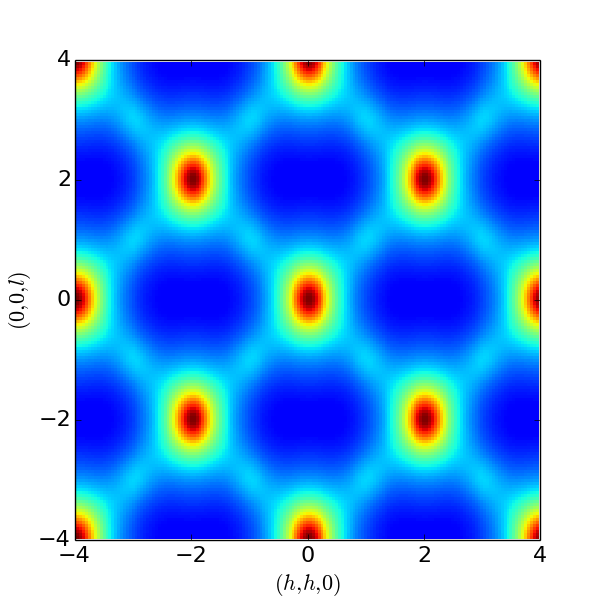}}%
\caption{(Color online)
Spin structure factor ${\mathcal{S}}(\mathbf{q})$
[Eq. (\ref{eq:strucfacdef)})]
 in the Ferromagnetic region of the phase diagram 
of $\mathcal{H}_{\sf breathing}$ [Eq. (\ref{eq:Hbreathing})]
for $\vartheta=\frac{-11 \pi}{16}$ (close to the non-breathing limit $J_A=J_B$)
and $\vartheta=\frac{-9 \pi}{16}$ (close to the isolated tetrahedron limit $J_A=0$)
[cf. Fig. \ref{fig:phasediagram}].
${\mathcal{S}}(\mathbf{q})$ is calculated from the Self Consistent Gaussian
Approximation (SCGA) described at the beginning of Section \ref{section:ferromagnet}.
With decreasing temperature the scattering sharpens around ${\bf q}=(0, 0, 0)$
and other ferromagnetic zone centers.
The sharpening is less pronounced at equal temperature for $\vartheta=\frac{-9 \pi}{16}$
because the growth of the correlation length at low temperatures is controlled by the
product $J_A J_B$ which vanishes in the isolated tetrahedron limit [Eqs. (\ref{eq:xieff})-(\ref{eq:lowTfm})].
}
\label{fig:FMSq}
\end{figure*}

In this Section we analyse the spin correlations where
the exchange coefficients $J_A$ and $J_B$ are
ferromagnetic. 
Here, and in the following two Sections where we analyse
the other regions of the phase diagram, we use the
Self Consistent Gaussian Approximation (SCGA) to calculate
the spin correlation functions.
This approach is known to be qualitatively succesful in treating the
spin correlations of classical vector spin models on the pyrochlore lattice 
\cite{canals01, isakov04, pickles08, conlon09, conlon10}.

In this treatment of the spin correlations, the ``hard spin'' constraint
that each vector spin have fixed length 
\begin{eqnarray}
{\bf S}_i \cdot {\bf S}_i=S^2 \qquad  \forall  \qquad i
\label{eq:hardspin}
\end{eqnarray}
is relaxed such that Eq. (\ref{eq:hardspin}) is satisfied only on average:
\begin{eqnarray}
\langle {\bf S}_i \cdot {\bf S}_i \rangle = S^2 \qquad  \forall  \qquad i
\label{eq:averageconstraint}
\end{eqnarray}

Eq. (\ref{eq:averageconstraint}) is enforced by means of a Lagrange multiplier $\lambda$.
Specifically, we write
\begin{eqnarray}
&&\beta \mathcal{H}_{\sf breathing} \to \beta \mathcal{H}_{\sf breathing}^{\sf SCGA}
\nonumber \\
&&\qquad =
\beta \mathcal{H}_{\sf breathing}+ (\lambda-\beta \epsilon_0) \sum_i {\bf S}_i^2.
\label{eq:HSCGA}
\end{eqnarray}
Here $\beta$ is the inverse temperature and
\begin{eqnarray}
\epsilon_0=\frac{E_0}{N S^2}
\end{eqnarray}
where $E_0$ is the ground state energy.
In Eq. (\ref{eq:HSCGA}) we have subtracted the constant term 
$\epsilon_0  \sum_i {\bf S}_i^2$ 
from the Hamiltonian
so that  $\mathcal{H}_{\sf breathing}^{\sf SCGA}$
is positive definite as long as $\lambda > 0$.
This step makes no difference to the results of our analysis, but is mathematically
convenient.

We define Fourier transformed
variables
\begin{eqnarray}
S_i^{\alpha}(\mathbf{q})=\sqrt{\frac{1}{N_{\sf uc}}}\sum_{{\bf r}_i} \exp(-i\mathbf{q} \cdot {\bf r}_i)S^{\alpha}(\mathbf{r}_i)
\label{eq:latticeFT}
\end{eqnarray}
where $\alpha=x,y,z$ indexes the spin components,
 $i=0,1,2,3$ indexes the four sublattices of the breathing pyrochlore lattice and $\sum_{{\bf r}_i}$
runs over all the sites ${\bf r}_i$ belonging to sublattice $i$.

\begin{figure}
\centering
\subfigure[$\vartheta=\frac{-11\pi}{16};  \ \ \frac{T}{\bar{J}}=5 ({\red \CIRCLE}), \
\frac{T}{\bar{J}}=4 ({\blue \blacksquare}),
\frac{T}{\bar{J}}=3 ({\green \blackdiamond}),
\frac{T}{\bar{J}}=2 ({\color{darkpurple} \blacktriangle})
 $]{%
\includegraphics[width=\columnwidth]{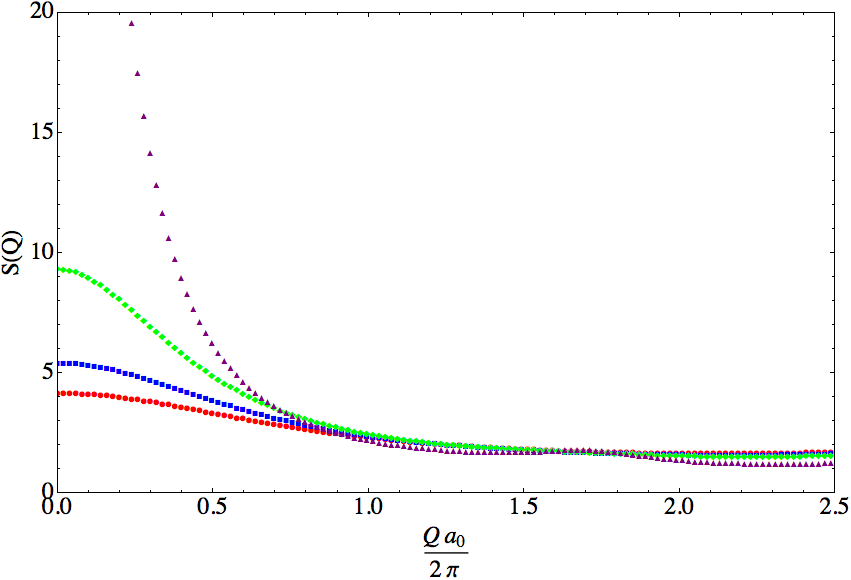}}%
\\
\subfigure[$\vartheta=\frac{-9\pi}{16};
\ \ \frac{T}{\bar{J}}=5 ({\red \CIRCLE}), \
\frac{T}{\bar{J}}=4 ({\blue \blacksquare}),
\frac{T}{\bar{J}}=3 ({\green \blackdiamond}),
\frac{T}{\bar{J}}=2 ({\color{darkpurple} \blacktriangle})$
]{%
\includegraphics[width=\columnwidth]{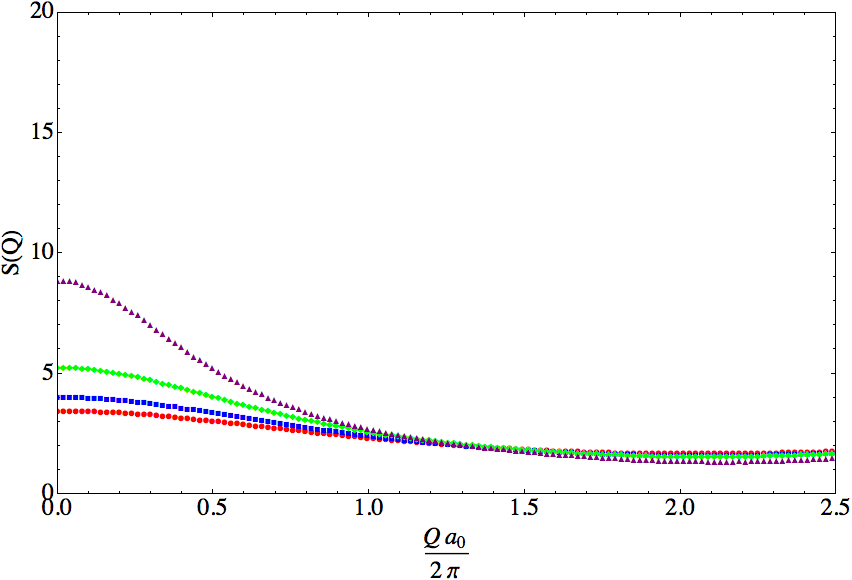}}%
\caption{(Color online)
Angle integrated structure factor  $\mathcal{S}_{\sf pow}(Q)$ [Eq. (\ref{eq:strucfacpow})] in the
ferromagnetic region of the phase diagram [Fig. \ref{fig:phasediagram}]
of \mbox{$\mathcal{H}_{\sf breathing}$ [Eq. (\ref{eq:Hbreathing})]}.
The structure factor is
 calculated from the SCGA as
described at the beginning of Section \ref{section:ferromagnet}.
Calculation is shown for $\vartheta=\frac{-11 \pi}{16}$
and
 $\vartheta=\frac{-9 \pi}{16}$.
Where the disparity between $J_A$ and $J_B$ is greater the feromagnetic
correlations build up more slowly with decreasing temperature.
}
\label{fig:FMpowder}
\end{figure}

Re-writing $\beta \mathcal{H}_{\sf breathing}^{\sf SCGA} $  [Eq. (\ref{eq:HSCGA})] in terms 
of $S_i^{\alpha}(\mathbf{q})$ [Eq. (\ref{eq:latticeFT})] we obtain
\begin{eqnarray}
 \beta \mathcal{H}_{\sf breathing}^{\sf SCGA}=
\frac{1}{2} \sum_{\mathbf{q}} \sum_{\alpha} \sum_{ij}
S^{\alpha}_i(-\mathbf{q}) \mathcal{M}_{ij} (\mathbf{q}) S^{\alpha}_j(\mathbf{q}).
\end{eqnarray}
The components of $4\times4$ matrix $\mathcal{M}_{ij} (\mathbf{q})$
are
\begin{eqnarray}
&& \mathcal{M}_{ij} (\mathbf{q}) \nonumber \\
&&\ \ =2 \delta_{ij}(\lambda-\beta \epsilon_0)+ \mathcal{A}_{ij}
\bigg[ J_A \exp\left(i \frac{d_A}{\sqrt{8}} \left({\bf e}_i-{\bf e}_j\right) \right)
\nonumber \\
&& \ \ 
 +J_B \exp\left(-i \frac{d_B}{\sqrt{8}} \left({\bf e}_i-{\bf e}_j\right) \right)
 \bigg] 
\end{eqnarray}
where
\begin{eqnarray}
&&\mathcal{A}=
\begin{pmatrix}
0 & 1 &1 & 1 \\
1 & 0 & 1 & 1 \\
1 & 1 & 0 & 1 \\
1 & 1 & 1 & 0
\end{pmatrix} \\
&&{\bf e}_0=(1, 1, 1)  \qquad
{\bf e}_1=(1, -1, -1)  \nonumber\\
&&{\bf e}_2=(-1, 1, -1) \quad
{\bf e}_3=(-1, -1, 1) 
\end{eqnarray}
and $d_A, d_B$ are the bond lengths on the $A$ and $B$
tetrahedra respectively [Fig. \ref{fig:lattice}].

The partition function within the SCGA is then
\begin{eqnarray}
\mathcal{Z}_{\sf SCGA}=\int \prod_{\bf q} \prod_{\alpha} dS^{\alpha}(\mathbf{q})
\exp\left(-\beta \mathcal{H}_{\sf breathing}^{\sf SCGA} \right).
\end{eqnarray}
Since $\beta \mathcal{H}_{\sf breathing}^{\sf SCGA}$ is Gaussian the spin correlation functions
within the SCGA are given by
\begin{eqnarray}
\langle S_i^{\mu}(-\mathbf{q}) S_j^{\nu} (\mathbf{q}) \rangle= \delta^{\mu \nu} \left[ \mathcal{M}(\mathbf{q})^{-1} \right]_{ij}.
\label{eq:SCGAcorr}
\end{eqnarray} 
The Lagrange multiplier $\lambda$ is then determined self consistently from 
Eqs. (\ref{eq:averageconstraint}) and (\ref{eq:SCGAcorr}).

The results of this analysis, for ferromagnetic exchange parameters $J_A, J_B<0$
are shown in \ref{fig:FMSq}  and Fig. \ref{fig:FMpowder}.

Fig. \ref{fig:FMSq} shows the spin structure factor
\begin{eqnarray}
\mathcal{S}(\mathbf{q})=\sum_{ij} \langle {\bf S}_i(-\mathbf{q}) \cdot {\bf S}_j(\mathbf{q}) \rangle
\label{eq:strucfacdef}
\end{eqnarray}
in the plane ${\bf q}=(h, h, l)$.
The structure factor is shown for two
temperatures $\frac{T}{\bar{J}}=5$ and 
$\frac{T}{\bar{J}}=2$ for $\theta=\frac{-11 \pi}{16}$ (close to the ``non-breathing'' limit
$J_A=J_B$) and $\theta=\frac{-9 \pi}{16}$ (close to the isolated tetrahedra limit $J_A=0$).
Fig. \ref{fig:FMpowder} shows the angle integral of Eq. (\ref{eq:strucfacdef}), 
for the same sets of exchange parameters and a range of temperatures, 
as would be measured in a neutron scattering
experiment on a powder sample
\begin{eqnarray}
&&\mathcal{S}_{\sf pow}(Q)= \frac{1}{4\pi} \int d\theta d\phi \sin(\theta)
\mathcal{S}\left(\bf{q} \right)  \\
&&{\bf q}=Q(\sin(\theta) \cos(\phi), \sin(\theta)\sin(\phi), \cos(\theta))
\label{eq:strucfacpow}
\end{eqnarray}

\begin{figure*}
\centering
\includegraphics[width=.8\textwidth]{colorbar.png}
\\
\subfigure[$\frac{T}{\bar{J}}=5, \vartheta=\frac{-\pi}{16}$]{%
\includegraphics[width=.33\textwidth]{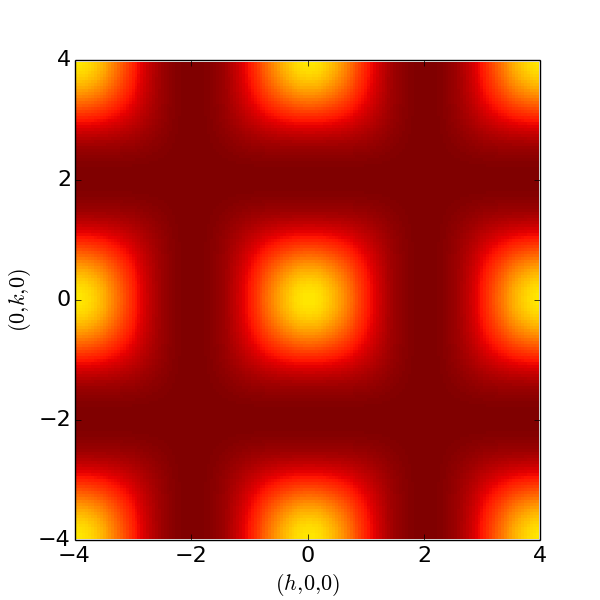}}%
\subfigure[$\frac{T}{\bar{J}}=5, \vartheta=\frac{-\pi}{4}$]{%
\includegraphics[width=.33\textwidth]{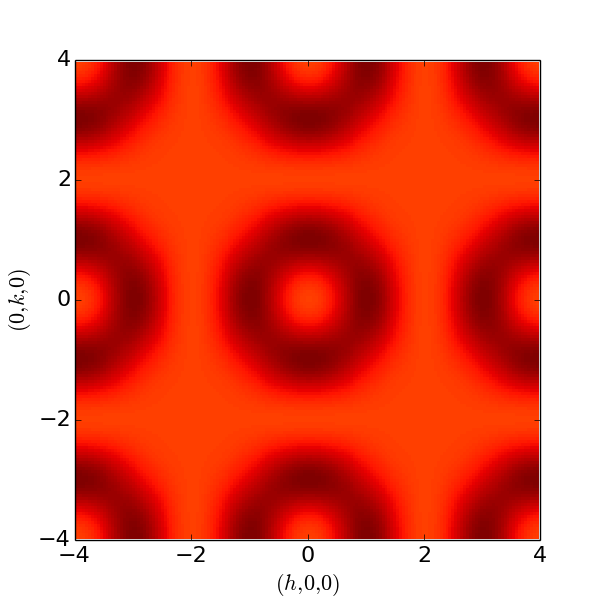}}%
\subfigure[$\frac{T}{\bar{J}}=5, \vartheta=\frac{-7\pi}{16}$]{%
\includegraphics[width=.33\textwidth]{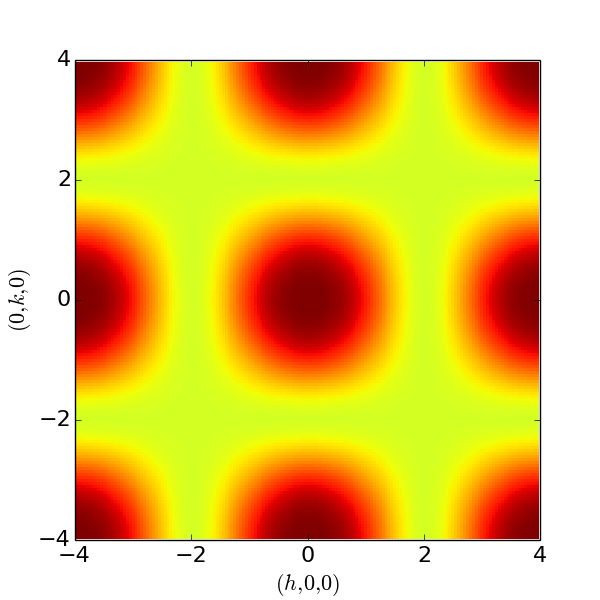}}%
\\
\subfigure[$\frac{T}{\bar{J}}=2, \vartheta=\frac{-\pi}{16}$]{%
\includegraphics[width=.33\textwidth]{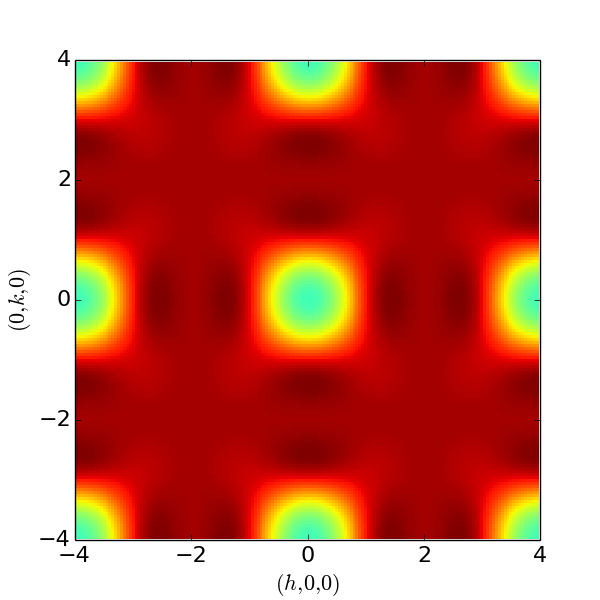}}%
\subfigure[$\frac{T}{\bar{J}}=2, \vartheta=\frac{-\pi}{4}$]{%
\includegraphics[width=.33\textwidth]{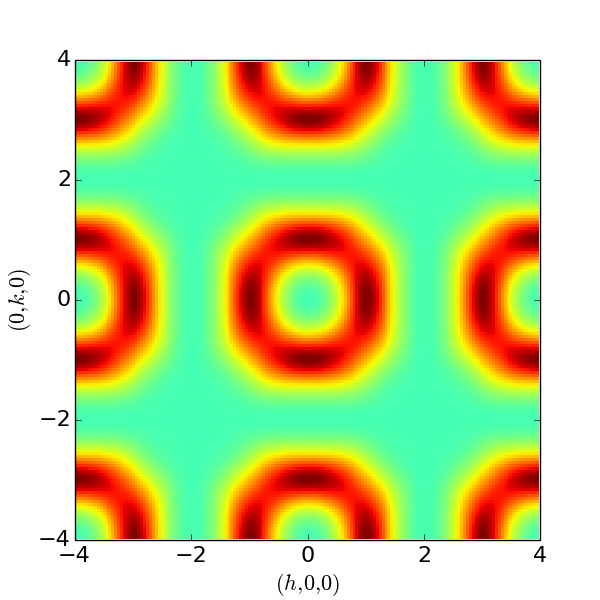}}%
\subfigure[$\frac{T}{\bar{J}}=2, \vartheta=\frac{-7\pi}{16}$]{%
\includegraphics[width=.33\textwidth]{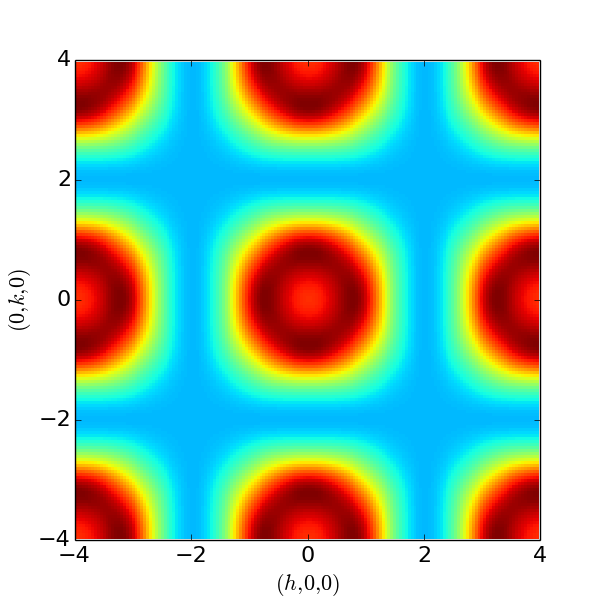}}%
\\
\subfigure[$\frac{T}{\bar{J}}=0.5, \vartheta=\frac{-\pi}{16}$]{%
\includegraphics[width=.33\textwidth]{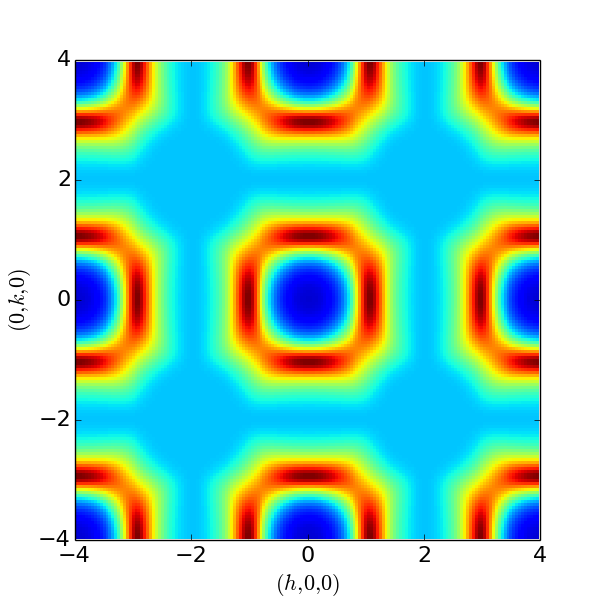}}%
\subfigure[$\frac{T}{\bar{J}}=0.5 \vartheta=\frac{-\pi}{4}$]{%
\includegraphics[width=.33\textwidth]{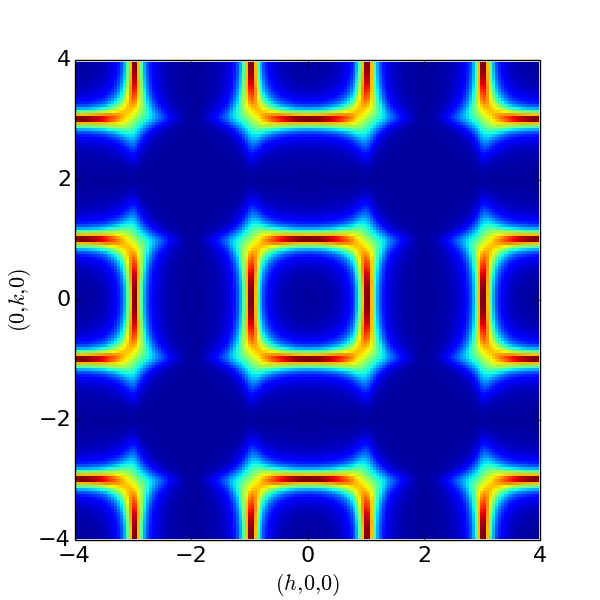}}%
\subfigure[$\frac{T}{\bar{J}}=0.5, \vartheta=\frac{-7\pi}{16}$]{%
\includegraphics[width=.33\textwidth]{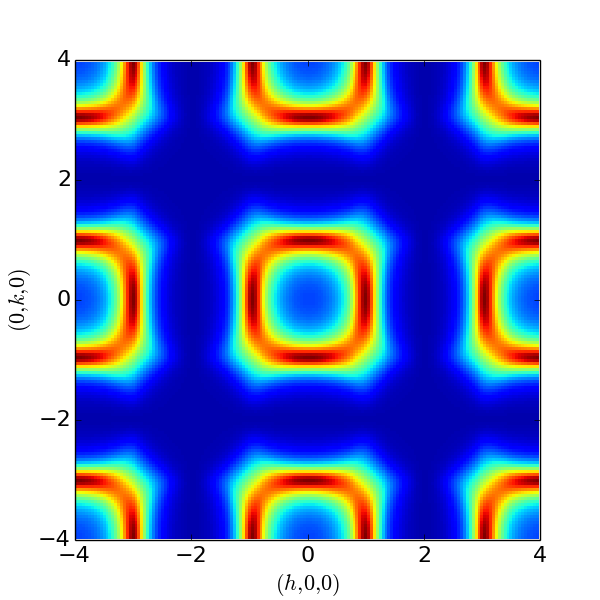}}%
\caption{(Color online)
Spin structure factor $\mathcal{S}(\mathbf{q})$ [Eq. (\ref{eq:strucfacdef})] for $\mathcal{H}_{\sf breathing}$
[Eq. (\ref{eq:Hbreathing})] in the region of the phase diagram [Fig. \ref{fig:phasediagram}]
 where $J_A$ and $J_B$ have opposite signs, in the plane ${\bf q}=(h, k, 0)$.
For all values of $\vartheta$ in this region the correlations evolve towards
a ``square ring'' pattern, which comes from interesecting lines of soft modes
with wavevectors of the form ${\bf q}=(1, \delta, 0)$
}
\label{fig:ringSq}
\end{figure*}

In both cases there is a sharpening of the scattering around into peaks around ${\bf q}=(0, 0, 0)$ 
(and other Brillouin zone centers associated with ferromagnetic Bragg peaks) as the temperature
approaches the ordering temperature $T_{\sf FM}$.
Within the SCGA this transition occurs when $\lambda(T)=0$. Since the SCGA includes
unphysical fluctuations which modulate the spin length, the resulting estimate of the transition
temperature is likely to be an underestimate of the true transition temperature.

The sharpening of the scattering is less pronounced (for equal temperature)
where the exchange parameters are closer to the isolated tetrahedra limit $J_A=0$.
In essence, where there is a large disparity between $J_A$ and $J_B$ the correlation length is limited by
the weaker of the two exchange parameters.

This can be understood quantitatively by considering the behaviour of the ferromagnetic correlation
length, obtained by expanding the SCGA calculation of Eq. (\ref{eq:strucfacdef}) around 
\mbox{${\bf q}=(0, 0, 0)$}.
For this purpose, and for the calculations in Sections \ref{section:orderbydisorder}
and \ref{section:coulombphase}, we will set the ratio of bond lengths on the tetrahedra  [Fig. \ref{fig:lattice}]
equal to unity
\begin{eqnarray}
\frac{d_A}{d_B}=1.
\end{eqnarray}
This assumption makes the analysis more transparent, and is not unphysical since in the real breathing pyrochlore materials $\frac{d_A}{d_B}$ is very close to unity, e.g. $0.95$ in LiInCr$_4$O$_8$ \cite{okamoto13}.
Where we come to compare our calculations to experiments on  LiInCr$_4$O$_8$ in Section \ref{section:discussion} we restore the true value of $\frac{d_A}{d_B}$.

In the region approaching ${\bf q}=(0, 0, 0)$ we have
\begin{eqnarray}
\mathcal{S}(\mathbf{q})=\frac{6}{\lambda(T)} \left( \frac{1}{1+\xi_{\sf eff}^2}\right)+ \mathcal{O}(q^4).
\end{eqnarray}
The square of the ferromagnetic correlation length $\xi_{\sf eff}^2$
may be written as the sum of two terms, with different behaviours approaching the ordering transition
\begin{eqnarray}
\xi_{\sf eff}^2=\xi_{0}^2+\xi_{\sf FM}^{2}.
\label{eq:xieff}
\end{eqnarray}
These two terms are
\begin{eqnarray}
&&\xi_{\sf 0}^2=\left(\frac{a_0}{4}\right)^2 \frac{\beta  |J_A+J_B|}{2(\lambda(T)+2\beta |J_A+J_B|) } \\
&&\xi_{\sf FM}^2=\left(\frac{a_0}{4}\right)^2 \frac{4 \beta^2 J_A J_B\lambda(T)}{\lambda(T)^2(\lambda(T)+2\beta |J_A+J_B|)} 
\end{eqnarray}
where $a_0$ is the cubic lattice parameter of the breathing pyrochlore lattice.

Approaching the ordering transition at $T=T_{\sf FM}$, we have
\begin{eqnarray}
\lim_{T \to T_{\sf FM}}\lambda(T)=0.
\end{eqnarray}
In this limit the correlation length $\xi_0$ saturates to a length scale of the order of the nearest neighbour
distance
\begin{eqnarray}
\lim_{\lambda \to 0}\xi_0(\lambda)^2=\frac{a_0^2}{8}.
\end{eqnarray}
Meanwhile the correlation length $\xi_{FM}$ diverges as $\frac{1}{\lambda}$
\begin{eqnarray}
\lim_{\lambda \to 0}\xi_{\sf FM}(\lambda)^2
=\left(\frac{a_0}{4}\right)^2 \frac{4 \beta^2 J_A J_B\lambda(T)}{\lambda(T)^22\beta |J_A+J_B|} 
\sim\frac{1}{\lambda}.
\end{eqnarray}
Therefore as $T\to T_{\sf FM}$ we have
\begin{eqnarray}
\xi_{\sf FM}^2 >> \xi_0^2 \implies \xi_{\sf eff} \approx \xi_{\sf FM}
\label{eq:lowTfm}
\end{eqnarray}

In the high temperature limit $\beta\to0$
we have
\begin{eqnarray}
\lim_{\beta \to 0}\xi_0^2=\left(\frac{a_0}{4}\right)^2 \frac{\beta |J_A+J_B|}{2 }  \sim \beta \\
\lim_{\beta \to 0}\xi_{\sf FM}^2=
\left(\frac{a_0}{4}\right)^2 \frac{4 \beta^2 J_A J_B}{\lambda(T)} 
\sim\beta^2
\end{eqnarray}
and so at high temperatures we have
\begin{eqnarray}
\xi_{\sf 0}^2 >> \xi_{\sf FM}^2 \implies \xi_{\sf eff} \approx \xi_{0}
\label{eq:highTfm}
\end{eqnarray}

Seen in this light, the two correlation lengths $\xi_0$ and $\xi_{\sf FM}$
are a quantitative measure of the extent to which the system may be thought of as
a set of isolated tetrahedra.
The crossover from the high temperature limit described by Eq. (\ref{eq:highTfm})
to the strongly correlated regime described by Eq. (\ref{eq:lowTfm}) is 
a thermal crossover from the physics of isolated tetrahedra to the physics of a long-range
correlated ferromagnet.
The relative magnitudes of $\xi_0$ and $\xi_{\sf FM}$ give a criterion for deciding 
whether a given
ferromagnetic breathing pyrochlore at a given temperature is in the ``isolated tetrahedron'' limit.

When either $J_A$ or $J_B$ vanishes, $\xi_{\sf FM}$ also vanishes for
all temperatures, since the system is then exactly described by a set of
isolated tetrahedra.

\section{$J_A$ and $J_B$ with opposite signs: dimensional reduction and order by disorder}
\label{section:orderbydisorder}

We now turn to discuss the case where $J_A$ and $J_B$ have opposite signs.
For concreteness we will take throughout this section
\begin{eqnarray}
J_A>0,  \quad J_B<0.
\end{eqnarray}

As discussed in Section \ref{section:phasediagram}, the ground state manifold in this
case has a degeneracy of $\mathcal{O}(L)$, with states in the ground state manifold
having associated wavevectors of the form ${\bf q}_{\delta}=(1, 0, \delta)$.
At some finite temperature $T_{\sf obd}$ the system will order
via the order--by--disorder mechanism.
The relationship between this model and the classical antiferromagnet on the FCC lattice
[Eq. (\ref{eq:HFCC})] suggests
that this will occur via a first-order phase transition \cite{gvozdikova05}
and that the resulting ordered state will have wavevector ${\bf q}_{\sf ord}=(1, 0, 0)$
(or those related by symmetry).

The development of these correlations 
in the plane \mbox{${\bf q}=(h, k, 0)$}
as a function of temperature and $\vartheta$ [Eq. (\ref{eq:thetadef})]
is shown in Fig. \ref{fig:ringSq}, for $\vartheta=-\frac{\pi}{16}, -\frac{\pi}{4}, -\frac{-7\pi}{16}$.
In all cases, the scattering develops 
with decreasing temperature
towards the ``square ring" pattern visible in Figs.
\ref{fig:ringSq} (g)-(i), which is formed from intersecting lines of scattering associated with
the soft modes at ${\bf q}_{\delta}=(1, \delta, 0)$, modulated by a zone dependent
form factor.
These bright lines of scattering signal an effective decoupling of neighbouring
$[001]$ planes- which is somewhat reminiscent of the dimensional
reduction observed in the rare-earth pyrochlore Yb$_2$Ti$_2$O$_7$
\cite{ross09, ross11-PRB84, yan-arXiv}.
In that system neighbouring $[111]$ planes become effectively decoupled 
and the consequence is rods of scattering along the $(h, h, h)$ directions of
reciprocal space.
Here, the rods are along the $(1, 0, h)$ directions and are modulated by a zone dependent
form factor resulting in the square ring structure.


\begin{figure}
\centering
\subfigure[$\vartheta=\frac{-\pi}{16}
\ \ \frac{T}{\bar{J}}=5 ({\red \CIRCLE}), \
\frac{T}{\bar{J}}=2 ({\blue \blacksquare}),
\frac{T}{\bar{J}}=1 ({\green \blackdiamond}),
\frac{T}{\bar{J}}=0.5 ({\color{darkpurple} \blacktriangle}),
\frac{T}{\bar{J}}=0.2 ({\yellow \blacktriangledown})
$]{%
\includegraphics[width=0.8\columnwidth]{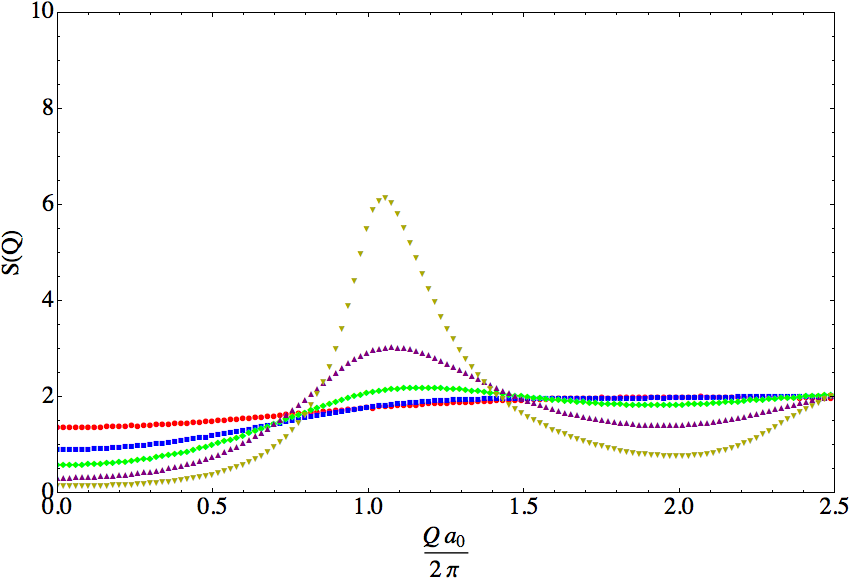}}%
\\
\subfigure[$\vartheta=\frac{-\pi}{4};
\ \ \frac{T}{\bar{J}}=5 ({\red \CIRCLE}), \
\frac{T}{\bar{J}}=2 ({\blue \blacksquare}),
\frac{T}{\bar{J}}=1 ({\green \blackdiamond}),
\frac{T}{\bar{J}}=0.5 ({\color{darkpurple} \blacktriangle}),
\frac{T}{\bar{J}}=0.2 ({\yellow \blacktriangledown})
$]{%
\includegraphics[width=0.8\columnwidth]{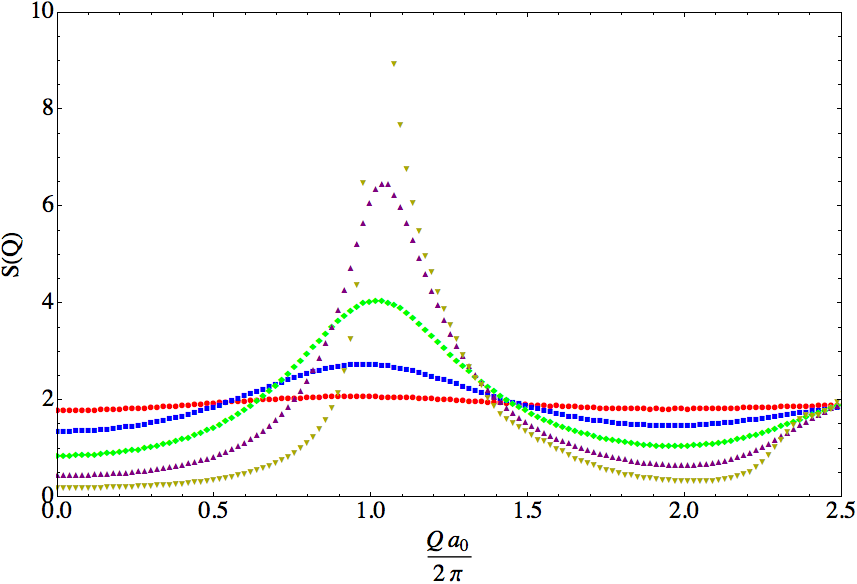}}%
\\
\subfigure[$\vartheta=\frac{-7\pi}{16};
\ \ \frac{T}{\bar{J}}=5 ({\red \CIRCLE}), \
\frac{T}{\bar{J}}=2 ({\blue \blacksquare}),
\frac{T}{\bar{J}}=1 ({\green \blackdiamond}),
\frac{T}{\bar{J}}=0.5 ({\color{darkpurple} \blacktriangle}),
\frac{T}{\bar{J}}=0.2 ({\yellow \blacktriangledown})
$]{%
\includegraphics[width=0.8\columnwidth]{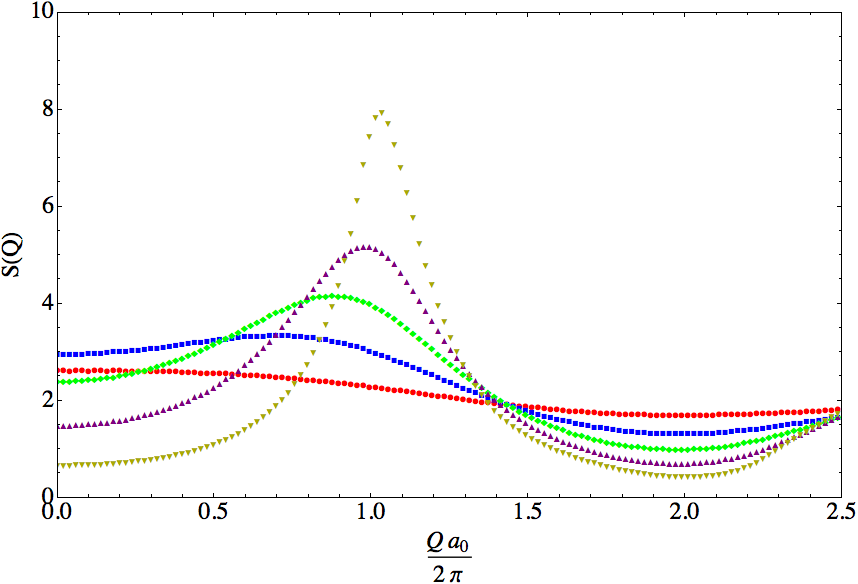}}%
\caption{(Color online)
Angle integrated structure factor  $\mathcal{S}_{\sf pow}(Q)$ [Eq. (\ref{eq:strucfacpow})] for $\mathcal{H}_{\sf breathing}$
[Eq.~(\ref{eq:Hbreathing})] in the region of the phase diagram [Fig. \ref{fig:phasediagram}]
where $J_A$ and $J_B$ have opposite signs.
At high temperatures, correlations are limited to single tetrahedra and 
are ferromagnetic or antiferromagnetic depending on the sign
of the dominant exchange paramter.
At low temperatures these correlations evolve into an increasingly sharp peak
near $\frac{Q a_0}{2 \pi}=1$, which is the result of antiferromagnetic correlations
between internally ferromagnetic tetrahedra.
The peak near $\frac{Q a_0}{2 \pi}=1$ comes from integrating the square ring
structure seen in Fig. \ref{fig:ringSq} (g)-(i).
}
\label{fig:ringpowder}
\end{figure}

In the case $\vartheta=\frac{-\pi}{4}$ [Figs. \ref{fig:ringSq} (b), (e), (h)],
where $|J_A|=|J_B|$ the scattering is only very weakly ${\bf q}$ dependent
at $\frac{T}{\bar{J}}=5$.
This reflects an effect by which the frustration arising from the opposite
exchange parameters, effectively reduces the temperature at which
appreciable ${\bf q}$ dependence in the correlations arises.
The ``square ring'' structure arises smoothly out of this, becoming increasingly
sharp as the temperature is lowered.

In the case  $\vartheta=\frac{-\pi}{16}$ [Figs. \ref{fig:ringSq} (a), (d), (g)],
the antiferromagnetic $J_A$ has a much larger magnitude than the ferromagnetic
$J_B$. 
The high temperature correlations therefore simply reflect antiferromagnetic
nearest-neighbour correlations on the $A$ tetrahedra, with a broad
minimum around ${\bf q}=(0, 0, 0)$ and a broad minimum 
at the neighbouring Brillouin zone centers (e.g.) ${\bf q}=(2, 0, 0)$.

At a finite temperature $T^{\sf AFM}_{\sf ring}$, given by the equation
\begin{eqnarray}
T^{\sf AFM}_{\sf ring}\lambda(T_{\sf AFM}^{\sf ring}) =\frac{-6 J_B(6J_A-2J_B)}{J_A+J_B}
\end{eqnarray}
the maximum at zone centers neighbouring
the first Brillouin zone  (e.g. ${\bf q}=(2, 0, 0)$) inverts into a
minimum, and a set of continuous set of maxima appear appear between
${\bf q}=(0, 0, 0)$ and the neighbouring Brillouin zone centers.
This set of maxima evolves in to the ``square ring'' feature seen at low temperature.
This temperature evolution of the correlations can be seen in angle integrated
scattering shown in Fig. \ref{fig:ringpowder}(a), where at broad maximum
centred at
$$
\frac{Q a_0}{2 \pi}=2
$$
at high temperature turns into a minimum,
with an increasingly sharp peak evolving
towards
$$
\frac{Q a_0}{2 \pi}\approx1.
$$
The peak in the powder integral always retains a finite
width and is shifted slightly to the right of 
$\frac{Q a_0}{2 \pi}=1$ because it has a contribution from all
wavevectors of the form ${\bf q}=(1, 0, \delta)$.

A similar picture describes the scattering when the
ferromagnetic exchange parameter dominates
the antiferromagnetic one.
This is the case when $\vartheta=\frac{-7\pi}{16}$
 [Figs. \ref{fig:ringSq} (c), (f), (i)].
Here the high temperature data has a broad maximum
at ${\bf q}=(0, 0, 0)$ reflecting ferromagnetic nearest
neighbour correlations on the $B$ tetrahedra.
This maximum inverts at a temperature given by
\begin{eqnarray}
T^{\sf FM}_{\sf ring} \lambda(T^{\sf FM}_{\sf ring})=\frac{J_A(6 J_B-2J_A)}{J_A+JB}
\end{eqnarray}
and a continuous set of maxima appear which
evolve towards  the ``square ring'' feature seen at low temperature.
The same evolution
can be seen in the powder integrated scattering  in Fig. \ref{fig:ringpowder}(c).

\begin{figure}
\centering
\includegraphics[width=.8\columnwidth]{colorbar.png}
\\
\subfigure[$\frac{T}{\bar{J}}=5, \vartheta=\frac{\pi}{16}$]{%
\includegraphics[width=.45\columnwidth]{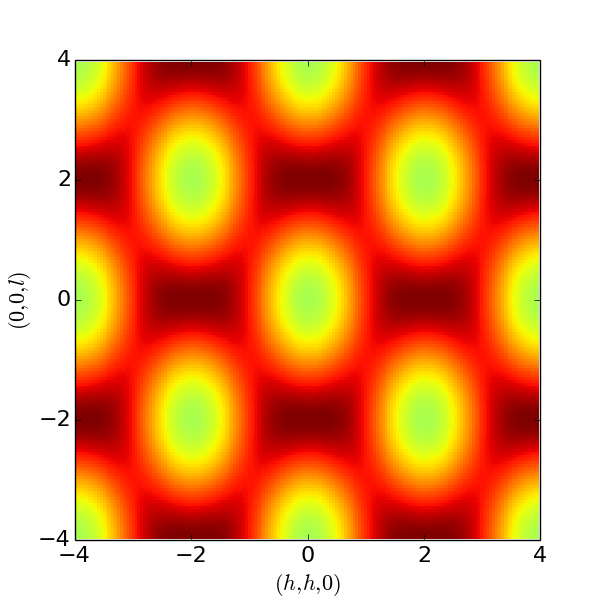}}%
\subfigure[$\frac{T}{\bar{J}}=5, \vartheta=\frac{3\pi}{16}$]{%
\includegraphics[width=.45\columnwidth]{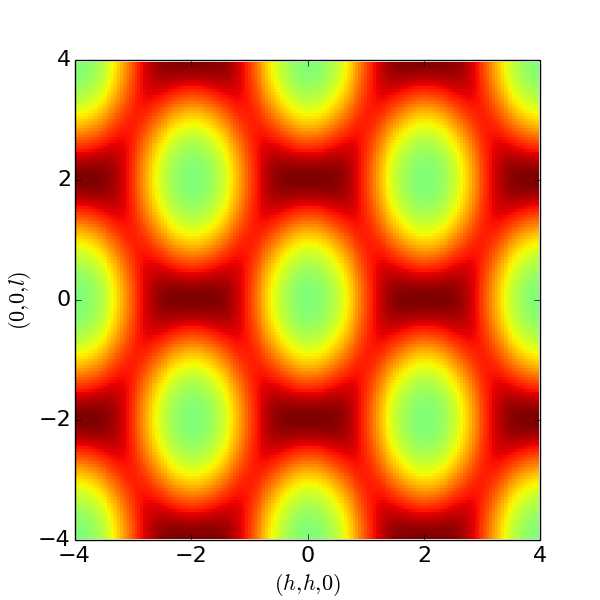}}%
\\
\subfigure[$\frac{T}{\bar{J}}=2, \vartheta=\frac{\pi}{16}$]{%
\includegraphics[width=.45\columnwidth]{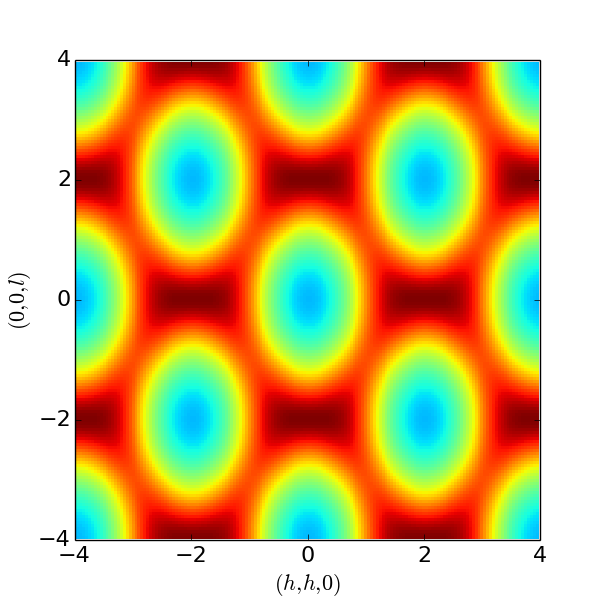}}%
\subfigure[$\frac{T}{\bar{J}}=2, \vartheta=\frac{3\pi}{16}$]{%
\includegraphics[width=.45\columnwidth]{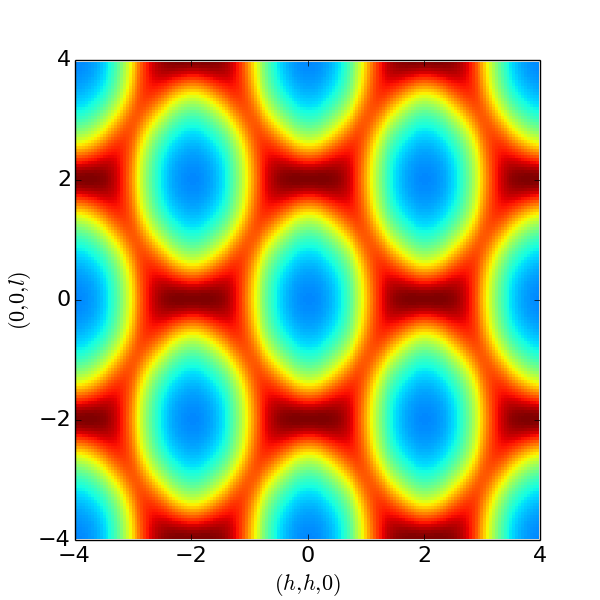}}%
\\
\subfigure[$\frac{T}{\bar{J}}=0.5, \vartheta=\frac{\pi}{16}$]{%
\includegraphics[width=.45\columnwidth]{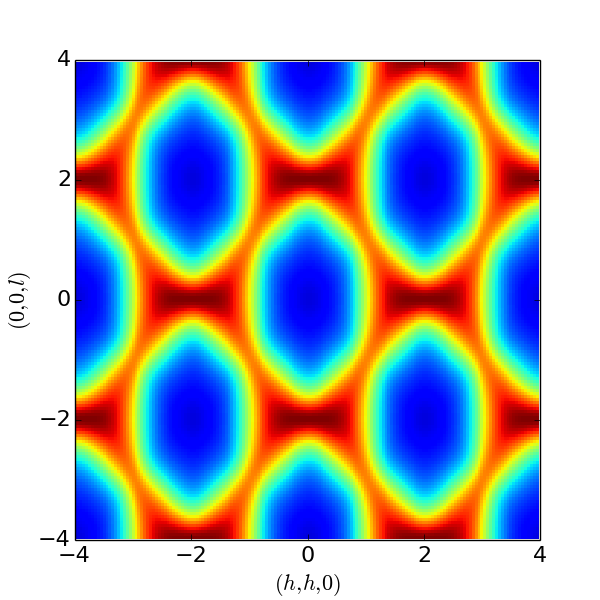}}%
\subfigure[$\frac{T}{\bar{J}}=0.5, \vartheta=\frac{3\pi}{16}$]{%
\includegraphics[width=.45\columnwidth]{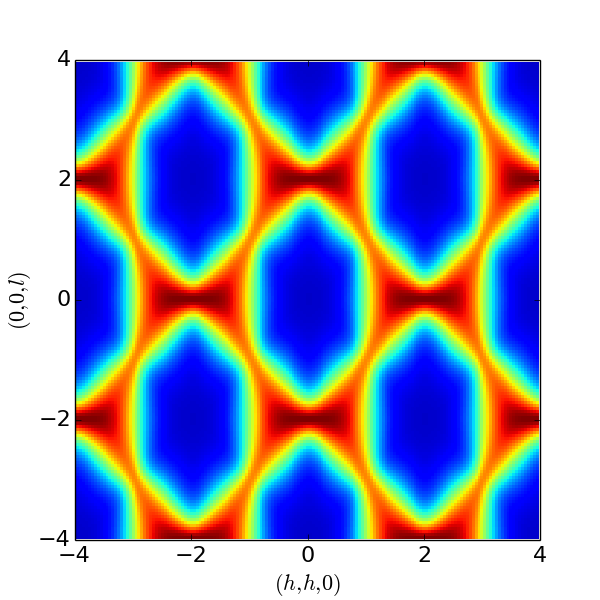}}%
\caption{(Color line)
Spin structure factor $\mathcal{S}(\mathbf{q})$ [Eq. (\ref{eq:strucfacdef})] in the antiferromagnetic (Coulomb phase)
region of the phase diagram [Fig. \ref{fig:phasediagram}]
of $\mathcal{H}_{\sf breathing}$ [Eq. (\ref{eq:Hbreathing})], shown for  $\vartheta=\frac{\pi}{16}$
and  $\vartheta=\frac{3\pi}{16}$.
With decreasing temperature, bow tie-like pinch point structures appear near
certain Brillouin zone centers, e.g. {\bf q}=(0, 0, 2).
As $T \to 0$ these become sharp singularities.
At finite temperature the singularity is cut off by a correlation length $\xi_{\parallel}$, 
which diverges as $T^{-1/2}$, signalling the onset of algebraic ($\sim r^{-3}$) correlations
in real space.
The divergent part of the correlation length is proportional to the product $J_A J_B$
and therefore the development of the correlations is slower when closer
to the isolated tetrahedron limit.
}
\label{fig:coulombSq}
\end{figure}

\begin{figure}
\centering
\subfigure[$\vartheta=\frac{\pi}{16};
\ \ \frac{T}{\bar{J}}=5 ({\red \CIRCLE}), \
\frac{T}{\bar{J}}=2 ({\blue \blacksquare}),
\frac{T}{\bar{J}}=1 ({\green \blackdiamond}),
\frac{T}{\bar{J}}=0.5 ({\color{darkpurple} \blacktriangle}),
\frac{T}{\bar{J}}=0.2 ({\yellow \blacktriangledown})$]{%
\includegraphics[width=\columnwidth]{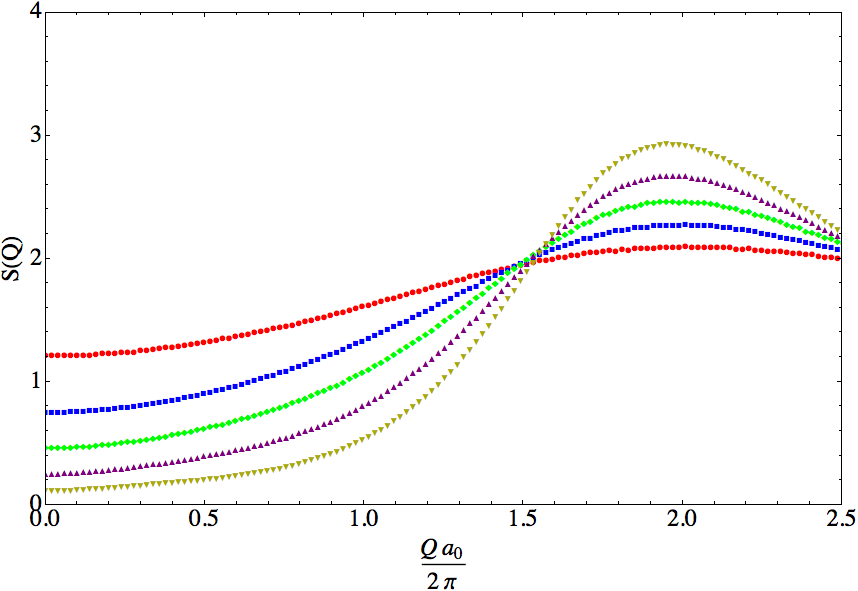}}%
\\
\subfigure[$\vartheta=\frac{3\pi}{16};
\ \ \frac{T}{\bar{J}}=5 ({\red \CIRCLE}), \
\frac{T}{\bar{J}}=2 ({\blue \blacksquare}),
\frac{T}{\bar{J}}=1 ({\green \blackdiamond}),
\frac{T}{\bar{J}}=0.5 ({\color{darkpurple} \blacktriangle}),
\frac{T}{\bar{J}}=0.2 ({\yellow \blacktriangledown})
$]{%
\includegraphics[width=\columnwidth]{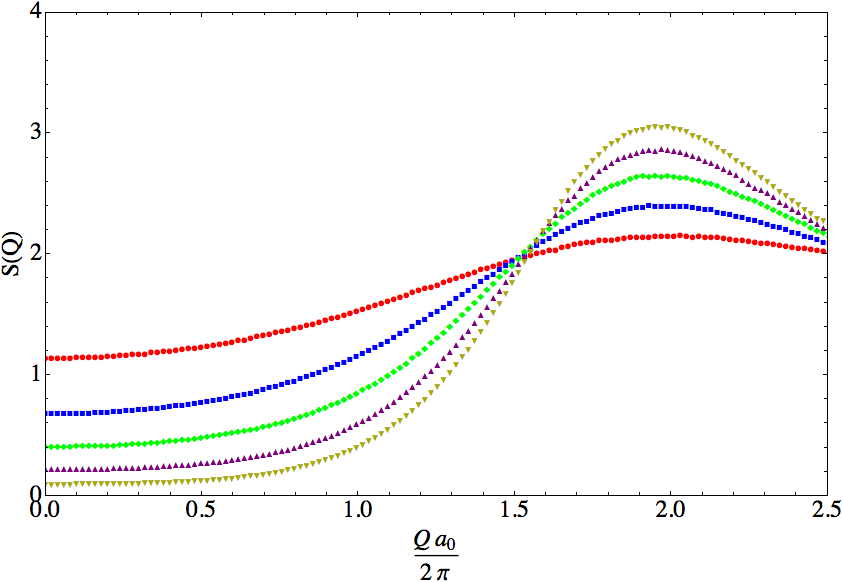}}%
\caption{(Color online)
Angle integrated structure factor $\mathcal{S}_{\sf pow}(\mathbf{q})$ [Eq. (\ref{eq:strucfacpow})] in the Coulombic spin liquid phase
of $\mathcal{H}_{\sf breathing}$ [Eq. (\ref{eq:Hbreathing})].
Correlations are
 shown for  $\vartheta=\frac{\pi}{16}$
and  $\vartheta=\frac{3\pi}{16}$ [cf. Fig. \ref{fig:phasediagram}].
The development of sharp pinch point features, as seen in Fig. \ref{fig:coulombSq}, 
is hidden by the angular integration and the powder structure factor shows a growing, broad
peak near $\frac{Q a_0}{2 \pi}$.
This is consistent with experimental observations on the breathing pyrochlore
LiInCr$_4$O$_8$ as discussed in Section \ref{section:discussion} and shown in Fig. \ref{fig:nilsenfit}.
}
\label{fig:coulombpowder}
\end{figure}

It is interesting to note that a ``square ring'' feature, rather similar to that seen
in Fig. \ref{fig:ringSq} has also been seen in reverse Monte Carlo reconstructions
of the scattering from the double perovskite material Ba$_2$YRuO$_6$
in which the Ru$^{5+}$ ions form an FCC lattice \cite{nilsen15}.
This material orders in to ${\bf q}=(1, 0, 0)$ antiferromagnetic
state shown in Fig. \ref{fig:FCC-AFM}(b) below $T=36$K.

\section{$J_A>0, \ J_B>0$: Coulombic spin liquid}
\label{section:coulombphase}

We now turn to discuss the 
spin correlations in the antiferromagnetic region of the phase diagram where
the system enters a Coulomb phase at low temperature.
These are shown for $\vartheta=\frac{\pi}{16}$ and
$\vartheta=\frac{3\pi}{16}$ in Fig. \ref{fig:coulombSq}
and Fig. \ref{fig:coulombpowder}.

The onset of the classical Coulomb phase is signalled by the sharpening
of bow-tie like ``pinch point'' features around reciprocal lattice
vectors such as ${\bf q}=(0, 0, 2)$.
In the $T\to0$ limit these features become sharp singularities \cite{isakov04, henley10}
but at finite tempeature they have a finite width in the longitudinal direction $\xi_{\parallel}^{-1}$.
The length scale $\xi_{\parallel}$ cuts off the algebraic 
real space correlations
at long distances, reverting
them to an exponential form for $r >> \xi_{\parallel}$. 
As $T\to0$, $ \xi_{\parallel}$ diverges as $T^{-1/2}$ and the correlations are algebraic out to 
infinite separation.

We may understand this crossover in the breathing pyrochlore case, by calculating $\xi_{\parallel}$
as a function of $J_A, J_B$ and $T$ within the SCGA. This is done by expanding
the SCGA prediction for $\mathcal{S}(\mathbf{q})$ around $(0, 0, 2)$.
We find
\begin{eqnarray}
\mathcal{S}((0, 0, 2+\tilde{q})) \approx
\frac{6}{\lambda(T)}\left(\frac{1}{1+\xi_{\parallel}^2 \tilde{q}^2} \right)+\mathcal{O}( \tilde{q}^4).
\end{eqnarray}

The square of the correlation length $\xi_{\parallel}^2$ may be written as a sum of two terms, one of
which reflects single tetrahedron correlations and dominates at high temperature the other
of which reflects the correlations of the Coulomb phase and dominates at low temperature
\begin{eqnarray}
&&\xi_{\parallel}^2=\xi_0^2+\xi_{\sf Coulomb}^2 \\
&&\xi_0^2=\left( \frac{a_0}{4}\right)^2 \left( \frac{2 \beta \lambda(T) (J_A+J_B)}{4 \lambda(T)^2+8 \beta \lambda(T) (J_A+J_B)}\right) \\
&&\xi_{\sf Coulomb}^2=
\left( \frac{a_0}{4}\right)^2
\left(
\frac{16 \beta^2 J_A J_B}{4 \lambda(T)^2 + 8 \beta \lambda(T) (J_A+J_B)}
\right) \nonumber \\
\end{eqnarray}

In the spin liquid region of the phase diagram
the Lagrange multiplier $\lambda(T)$ tends to a finite value at both high and low temperature.
Thus, in the low temperature limit we have
\begin{eqnarray}
&&
\lim_{T\to0}
\xi_0^2=\left( \frac{a_0}{8}\right)^2 
 \\
&&
\lim_{T\to0}
\xi_{\sf Coulomb}^2=
\left( \frac{a_0}{4}\right)^2
\left(
\frac{2 \beta J_A J_B}{ \lambda(T) (J_A+J_B)}
\right)\sim\frac{1}{T} \nonumber \\
\\ 
&&
\xi_0^2<<\xi_{\sf Coulomb}^2\implies \xi_{\parallel} \approx \xi_{\sf Coulomb} \sim T^{-1/2}
\end{eqnarray}
and the correlation length diverges as $T^{-1/2}$ as expected for a Coulomb phase.
In the case $J_A=J_B$ these expressions reproduce the results of Conlon and Chalker
\cite{conlon10}.

In the high temperature limit we have
\begin{eqnarray}
&&
\lim_{\beta\to0}
\xi_0^2=\left( \frac{a_0}{4}\right)^2 \left( \frac{2 \beta(J_A+J_B)}{4 \lambda(T)}\right)
\sim \beta
 \\
&&
\lim_{\beta\to0}
\xi_{\sf Coulomb}^2=
\left( \frac{a_0}{4}\right)^2
\left(
\frac{16 \beta^2 J_A J_B}{4 \lambda(T)^2 }
\right)
\sim \beta^2 \\
&&
\xi_{\sf Coulomb}^2<<\xi_0^2\implies \xi_{\parallel} \approx \xi_{\sf 0} 
\end{eqnarray}
and the correlations are essentially described by a model of isolated tetrahedra.
Comparison of the length scales $\xi_{\sf Coulomb}$ and $\xi_0$ gives a criterion
to define the thermal crossover from isolated tetrahedron correlations to
isolated Coulomb phase correlations in an antiferromagnetic 
breathing pyrochlore.

In the powder structure factor shown in Fig. \ref{fig:coulombpowder} the sharpness of the pinch point
features
is hidden by the angular integration [Eq. (\ref{eq:strucfacpow})] and the scattering simply shows a 
broad maximum around
\begin{eqnarray}
\frac{Qa_0}{2 \pi}=2.
\end{eqnarray}
This maximum grows in intensity and becomes sharper with decreasing temperature but 
remains always rather broad due to the angular integration.
This is consistent with neutron scattering experiments on LiInCr$_4$O$_8$, as
will be discussed further in Section \ref{section:discussion}.

\section{Discussion}
\label{section:discussion}

\begin{figure}
\centering
\includegraphics[width=\columnwidth]{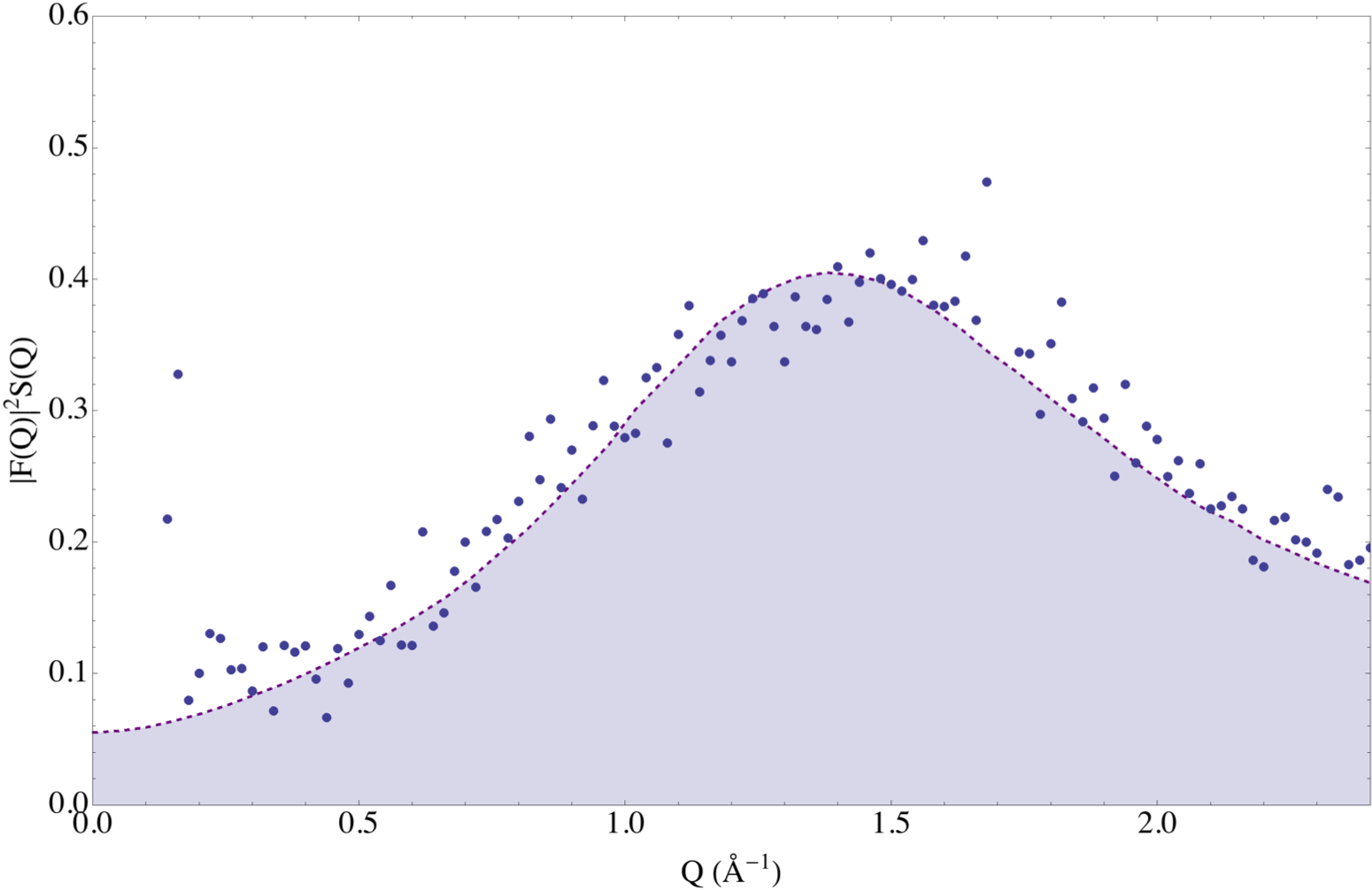}%
\caption{(Color online) Comparison of SCGA calculations
with experimental neutron scattering data \cite{okamoto15} 
for a powder sample of
for LiInCr$_4$O$_8$ at $T=30 K$.
The lattice constant $a_0=8.42 \AA $ and the bond lengths
$d_A=2.903 \AA$ and $d_B=3.052 \AA$ are set to their
experimental values \cite{okamoto13}. 
The exchange constants
are taken from an estimate by Okamoto {\it et al.} based on an empirical realtionship
between the bond lengths and the exchange parameters in Cr spinels \cite{okamoto13}:
 $J_A=60$K and $J_B$=6K.
The SCGA calculation of the scattering has been multiplied by the form
factor for Cr$^{3+}$ ions $|F(Q)|^2$.
There is one adjustable parameter in the comparison which
is a multiplicative scale factor setting the overall normalisation.
The agreement is very good, which suggests that the estimated
exchange parameters are approximately correct in this temperature range
and underlines the validity of the SCGA description in the paramagnetic,
cubic symmetry phase of  LiInCr$_4$O$_8$.
}
\label{fig:nilsenfit}
\end{figure}

In this article we have considered the minimal model for
for magnetism on the breathing pyrochlore lattice [Eq. (\ref{eq:Hbreathing})].
We have seen that the ground state manifold depends only on the signs of the
two exchange couplings, associated with the two species of tetrahedra [Section 
\ref{section:phasediagram}].
The classical ground state phase diagram 
[Fig. \ref{fig:phasediagram}] contains a ferromagnetic phase (for $J_A<0, J_B<0$)
a disordered Coulomb phase  (for $J_A>0, J_B>0$) and a phase where
an $\mathcal{O}(L)$ degeneracy is lifted by the order--by--disorder mechanism 
(for $J_A>0, J_B<0$ or vice versa).
The temperature development of the spin correlations in each case is calculated
using the Self Consistent Gaussian Approximation (SCGA) \cite{canals01, pickles08, conlon09, conlon10}
and is discussed for each region of the phase diagram in Sections \ref{section:ferromagnet}-
\ref{section:coulombphase}.

At this point it is interesting to compare our calculations with the results of the recent neutron scattering
experiments by Okamoto {\it et al.} \cite{okamoto15}.
These authors measured the neutron scattering response of powder samples of breathing
pyrochlores LiIn$_x$Ga$_{1-x}$Cr$_4$O$_8$.
Their neutron scattering data for $x=1$ (i.e. for LiInCr$_4$O$_8$)
at $T=30$K, are shown in Fig. \ref{fig:nilsenfit}.

In Ref. \citen{okamoto13} the authors provided an estimate of the
exchange parameters for LiInCr$_4$O$_8$,
based on an empirical relationship between the exchange parameters
and the bond distances,
obtaining  
\begin{eqnarray}
J_A=60\text{K}, \quad J_B=6 \text{K}.
\label{eq:okamotoparams}
\end{eqnarray}

We have compared  the results of our SCGA calculation using this parameter set 
with the neutron scattering data from Ref. \citen{okamoto15}.
The result is 
 shown in Fig. \ref{fig:nilsenfit}.
The comparison shows very good agreement, with one adjustable parameter which
is a multiplicative scale factor setting the overall normalisation.
This suggests that the exchange parameters proposed in Ref.~\citen{okamoto13} [Eq. (\ref{eq:okamotoparams})]
are at least approximately correct, at temperatures above the structural
transition and underlines the validity of the description presented here.
If we fix the overall energy scale $\bar{J}=\sqrt{J_A^2+J_B^2}$ [Eq.~(\ref{eq:thetadef})]
and vary the ratio $\frac{J_B}{J_A}$ we find that reasonably good agreement is obtained
for the range of values 
\begin{eqnarray}
0.05 \lesssim \frac{J_B}{J_A} \lesssim 0.15.
\end{eqnarray}

Rather surprisingly, we are unable to obtain a convincing fit to the higher temperature
data at $T=150$K.
As noted by the authors of Ref. \citen{okamoto15},
the experimental scattering at  $T=150$K is simply form-factor
like, suggesting vanishing spin correlations, 
whereas our analysis predicts measurable
nearest-neighbour spin correlations at $T=150$K for the estimated
$J_A=60$K and $J_B=6$K.
If we allow $J_A$ and $J_B$ to vary, we are unable to obtain any parameter set which simultaneously
describes the data at both temperatures.

This may be an indication that coupling to the lattice gives rise to an effectively temperature
dependent exchange energy scale. 
This seems plausible since the dominant exchange pathway is direct exchange between
Cr$^{3+}$ ions, which is highly sensitive to the Cr-Cr distance \cite{yaresko08}.

In conclusion, the recently synthesized breathing pyrochlore materials
offer new opportunities in the study of frustrated magnetism, playing
host to spin--liquid and order--by--disorder physics.
It may be particularly interesting for future experimental studies to 
investigate sulfide and selenide breathing pyrochlores \cite{pinch70}
which could potentially realise the case of opposite exchange parameters
discussed in Section \ref{section:orderbydisorder}, 
which exhibits an unusual structure of spin correlations and
order--by--disorder.

The determination of the full phase diagram of \mbox{$\mathcal{H_{\sf breathing}}$
~[Eq. (\ref{eq:Hbreathing})]} in the presence of quantum
fluctuations remains an important and interesting open question.
Moreover, experimental results on the breathing pyrochlores LiInCr$_4$O$_8$
and LiGaCr$_4$O$_8$ make it evident that the effect of coupling
to the lattice is an important direction for future work \cite{okamoto13, tanaka14, okamoto15, nilsen15-1}.
We expect that as in the ``non-breathing'' case
\cite{yamashita00, tchernyshyov02, penc04}, the physics uncovered by a treatment of the
spin-lattice coupling will be very rich indeed.

\section*{Acknowledgements}
We thank Zenji Hiroi,
G{\o}ran Nilsen and Yoshihiko Okamoto
for discussions about experiments,
for  helpful comments on the
manuscript and for   
sharing the neutron scattering data from Ref. \citen{okamoto15}, shown in Fig.~\ref{fig:nilsenfit}.
This work was supported by 
the Theory of Quantum Matter unit of the
Okinawa Institute of Science and Technology Graduate University



\end{document}